\documentclass[usenatbib]{mnras}

\usepackage[T1]{fontenc}
\usepackage{ae,aecompl}
\usepackage{hyperref} 
\usepackage{longtable} 
\usepackage{threeparttablex}  
\usepackage{array}  

\usepackage{graphicx}	
\usepackage{amsmath}	
\usepackage{amssymb}	

\newcommand{\bc}{\begin{center}}
\newcommand{\ec}{\end{center}}

\newcommand{\Mpc}            {\,{\rm Mpc}}
\newcommand{\Msun}           {\,{\rm M}_\odot}

\newcommand{\kms}            {\,{\rm km}\,\,{\rm s}^{-1}}

\newcommand{\Mstr}          {M_{\rm str}}
\newcommand{\Vmax}          {V_{\rm max}}
\newcommand{\acc}           {{\, {\rm m}\,{\rm s}^{-2}}}

\title[Tidal stripping and the structure of LG dwarfs] {Tidal
  stripping and the structure of dwarf galaxies in the
  Local Group}
\author[A. Fattahi et al. ]{
\parbox[t]{\textwidth}{ Azadeh Fattahi$^{1,2}$\thanks{Email:
    azadeh.fattahi-savadjani@durham.ac.uk}, Julio F. Navarro$^{2,3}$, Carlos S. Frenk$^{1}$, Kyle
  A. Oman$^{4}$, \\
  Till Sawala$^{5}$ and Matthieu Schaller$^{1}$ } \\ \\
\parbox[t]{\textwidth}{
  $^{1}$Institute for Computational Cosmology, Department of Physics,
  University of Durham, South Road, Durham DH1 3LE, United Kingdom\\
  $^2$Department of Physics and Astronomy, University of Victoria, PO Box
  3055 STN CSC, Victoria, BC, V8W 3P6, Canada\\
  $^3$Senior CIfAR Fellow.\\
  $^4$Kapteyn Astronomical Institute, University of Groningen,
  Postbus 800, NL-9700 AV Groningen, The Netherlands\\
  $^{5}$Department of Physics, University of Helsinki, Gustaf
  H\"allstr\"omin katu 2a, FI-00014 Helsinki, Finland \\
} }

\date{Accepted XXX. Received YYY; in original form ZZZ}

\pubyear{2017}

\begin{document}
\label{firstpage}
\pagerange{\pageref{firstpage}--\pageref{lastpage}}
\maketitle

\begin{abstract} 
  The shallow faint-end slope of the galaxy mass function is usually
  reproduced in $\Lambda$CDM galaxy formation models by assuming that
  the fraction of baryons that turns into stars drops steeply with
  decreasing halo mass and essentially vanishes in haloes with maximum
  circular velocities $V_{\rm max}<20$--$30 \kms$. Dark
  matter-dominated dwarfs should therefore have characteristic
  velocities of about that value, unless they are small enough to
  probe only the rising part of the halo circular velocity curve
  (i.e., half-mass radii, $r_{1/2}\ll 1$ kpc). Many dwarfs have
  properties in disagreement with this prediction: they are large
  enough to probe their halo $V_{\rm max}$ but their characteristic
  velocities are well below $20 \kms$. These `cold faint giants' (an
  extreme example is the recently discovered Crater~2 Milky Way
  satellite) can only be reconciled with our $\Lambda$CDM models if
  they are the remnants of once massive objects heavily affected by
  tidal stripping. We examine this possibility using the APOSTLE
  cosmological hydrodynamical simulations of the Local Group. Assuming
  that low velocity dispersion satellites have been affected by
  stripping, we infer their progenitor masses, radii, and velocity
  dispersions, and find them in remarkable agreement with those of
  isolated dwarfs. Tidal stripping also explains the large scatter in
  the mass discrepancy-acceleration relation in the dwarf galaxy
  regime: tides remove preferentially dark matter from satellite
  galaxies, lowering their accelerations below the
  $a_{\rm min}\sim 10^{-11} \acc$ minimum expected for isolated
  dwarfs. In many cases, the resulting velocity dispersions are
  inconsistent with the predictions from Modified Newtonian Dynamics,
  a result that poses a possibly insurmountable challenge to that
  scenario.
\end{abstract}


\section{Introduction}
\label{SecIntro}

The standard model of cosmology, Lambda Cold Dark Matter
($\Lambda$CDM), makes clear predictions for the dark halo mass
function once the cosmological parameters are specified
\citep{Jenkins2001,Tinker2008,Angulo2012}. At the low mass end, this
is much steeper than the faint end of the galaxy stellar mass
function, an observation that precludes a simple, linear relation
between galaxy and halo masses at the faint end. The difference can
be resolved if galaxies fail to form in haloes below some `threshold'
mass; this confines galaxies to relatively massive haloes, preventing
the formation of large numbers of faint dwarfs and reconciling the
faint-end slope of the galaxy luminosity function with the predictions
of $\Lambda$CDM \citep[see, e.g.,][and references therein]{White1991,Benson2003}.
 
This is not simply an ad-hoc solution. QSO studies have long indicated
that the Universe reionized soon after the first stars and galaxies
formed \citep[$z_{\rm reion}\lesssim 8$; see, e.g.,][]{Fan2006b}, an
event that heated the intergalactic medium to the ionization energy of
hydrogen, evaporating it away from low-mass haloes and proto-haloes,
especially from those that had not yet been able to collapse. In
slightly more massive haloes, where gas is able to collapse, vigorous
winds powered by the energy of the first supernovae expel the
remaining gas. These processes thus provide a natural explanation for
the steeply declining galaxy formation efficiency with decreasing halo
mass required to match the faint end of the galaxy stellar mass
function. Cosmological galaxy formation simulations, such as those
from the APOSTLE/EAGLE \citep{Schaye2015,Sawala2016b} or Illustris
projects \citep{Vogelsberger2014} rely heavily on this mechanism to
explain not only the faint-end of the luminosity function, but also
the abundance of Galactic satellites, their stellar mass distribution,
and their dark matter content \citep[see; e.g.,][]{Sawala2016a}.

Simulations like APOSTLE\footnote{APOSTLE: A Project Of Simulating The
  Local Environment.} predict a tight correlation between galaxy mass
and halo mass; given the stellar mass of a galaxy, $\Mstr$, its halo
mass is constrained to better than $\sim~15$ per cent in the dwarf
galaxy regime, defined hereafter as $\Mstr <10^9\, M_\odot$. Because
of the steep mass dependence of the galaxy formation efficiency in
this mass range the converse is not true: at a given halo mass
galaxies scatter over decades in stellar mass, in
  agreement with the latest semi-analytic models of galaxy formation
  \citep{Moster2017}. This is especially true of `faint dwarfs',
defined as those fainter than $\Mstr \sim 10^7\, M_\odot$ (about the
mass of the Fornax dwarf spheroidal), which are all expected to form
in haloes of similar mass, or, more specifically, haloes with maximum
circular velocity in the range
$20 \lesssim V_{\rm max}/{\rm km\, s}^{-1} \lesssim 30$ \citep[see;
e.g.,][]{Okamoto2009,Sawala2016b,Oman2016}.

This observation has a couple of important corollaries. One is that,
since the dark mass profile of CDM haloes is well constrained
\citep[][hereafter NFW]{Navarro1996,Navarro1997}, the dark matter
content of faint dwarfs should depend tightly on
their size: physically larger galaxies are expected to enclose more dark
matter and have, consequently, higher velocity dispersions. A second
corollary is that galaxies large enough to sample radii close to
$r_{\rm max}$, where the halo circular velocity reaches its maximum
value, $V_{\rm max}$, should all have similar characteristic circular velocities of
order $20$--$30 \kms$ , reflecting the narrow range of their parent halo
masses. For this velocity range, $r_{\rm max}$ is expected to be of
order $\sim 3$--$6$~kpc, and faint dwarfs as large as $\sim 1$ kpc should have
circular velocities well above $\sim~15 \kms$.

At first glance, these corollaries seem inconsistent with the
observational evidence. Indeed, there is little correlation between
velocity dispersion and size in existing faint dwarf samples, and
there are a number of dwarfs that, although large enough to sample
radii close to $r_{\rm max}$, still have velocity dispersions well
below $\sim 20 \kms$. A prime example is the recently discovered
Crater 2 dwarf spheroidal \citep{Torrealba2016}, termed a `cold faint
giant' for its large size (projected half-mass radius $r_{1/2}\sim 1$
kpc), low stellar mass ($\Mstr \sim 10^{5} \,M_\odot$) and small
velocity dispersion
\citep[$\sigma_{\rm los}\sim 3 \kms$,][]{Caldwell2017}.  The basic
disagreement between the relatively large velocities expected for
dwarfs and the low values actually measured is at the root of a number
of `challenges' to $\Lambda$CDM on small scales identified in recent
years \citep[see, e.g., the recent reviews by][]{DelPopolo2017,Bullock2017}.

Before rushing to conclude that these problems signal the need for a
radical change in the cold dark matter paradigm, it is important to
recall that the corollaries listed above rest on two important
assumptions: one is that (i) the assembly of a dwarf does not change
appreciably the dark matter density profile, and another is that
(ii) dwarfs have evolved in isolation and have not been subject to the
effects of external tides, which may in principle substantially alter their
dark matter and stellar content.

The first issue has been heavily debated in the literature, where,
depending on the algorithmic choice made for star formation and
feedback, simulations show that the baryonic assembly of the galaxy
can in principle reduce the central density of dark matter haloes and
create `cores'
\citep{Navarro1996b,Read2005,Mashchenko2006,Governato2012,Pontzen2014,Onorbe2015},
or not \citep{Schaller2015b,Oman2015,Vogelsberger2014}. Consensus has
yet to be reached on this issue but we shall use for our discussion
simulations that support the more conservative view that faint dwarfs
are unable to modify substantially their dark haloes. If
baryon-induced cores are indeed present in this mass range (and are large
enough to be relevant), they would only help to ease the difficulties
that arise when contrasting theoretical $\Lambda$CDM expectations with
observation.

The second issue is also important, since much of what is known about
the faintest galaxies in the Universe has been learned from samples
collected in the Local Group (LG), and therefore include satellites of
the Milky Way (MW) and Andromeda (M31), which may have been affected
by the tidal field of their hosts.  It is therefore important to
consider in detail the potential effect of tidal stripping on the
structural properties of satellites and their relation to isolated
dwarfs. Tides have been long been argued to play a
  critical role in determining the mass and structure of satellites
  \citep[see, e.g.,][and references
  therein]{Mayer2001,Kravtsov2004b,DOnghia2009,Kazantzidis2011,Tomozeiu2016,Frings2017}.  We
  address this issue here using a combination of direct cosmological
  hydrodynamical simulations complemented with the tidal stripping
  models of \citet[][hereafter PNM08]{Penarrubia2008b} and
  \citet[][E15]{Errani2015}, which parametrise the effect of tidal
  stripping in a particularly simple way directly applicable to
  observed dwarfs. We are thus able to track tidally-induced changes
  even for very faint dwarf satellites, where cosmological simulations
  are inevitably compromised by numerical limitations.

This paper is organized as follows. Sec.~\ref{SecObs} describes the
observational sample we use in this study, and the procedure we use to
estimate their dark matter content from their half-light radii and
velocity dispersions. The APOSTLE hydrodynamical simulations are
introduced in Sec.~\ref{SecSims}, followed by a discussion of the
galaxy mass-halo mass relation in Sec.~\ref{SecMstarVc}. The effects
of tidal stripping are discussed in in Sec.~\ref{SecTidStrip}; their
implications for the mass discrepancy-acceleration relation (MDAR) are
discussed in Sec~\ref{SecMDAR}, and for Modified Newtonian Dynamics
(MOND) in Sec~\ref{SecMOND}. We summarize our main conclusions in
Sec.~\ref{SecConc}.

\section{Observational data}
\label{SecObs}

\subsection{Dynamical masses}
\label{Sec DynMass}

The total mass within the half-light radius of velocity
dispersion-supported stellar systems, such as dwarf spheroidals
(dSphs), can be robustly estimated for systems that are close to
equilibrium, reasonably spherical in shape, and with constant or
slowly varying velocity dispersion profiles
\citep[e.g.,][]{Walker2009d}. \citet{Wolf2010}, in particular, show
that the enclosed mass within the 3D (deprojected) half-light radius
($r_{1/2}$) may be approximated by
\begin{equation}
 M_{1/2}=3\,G^{-1}\,\sigma_{\rm los}^2\,r_{1/2},
\label{eqW10}
\end{equation} 
where $\sigma_{\rm los}$ is the luminosity-weighted
line-of-sight velocity dispersion of the stars and $r_{1/2}$ has been
derived from the (projected) effective radius, $R_{\rm eff}$, using
$r_{1/2}=(4/3)R_{\rm eff}$.

We adopt Eq.~\ref{eqW10} to estimate $M_{1/2}$ for all dwarf galaxies
in the LG with measured velocity dispersion and effective radius. As
is customary, we use the circular velocity at $r_{1/2}$ as a measure
of mass, instead of $M_{1/2}$:
\begin{equation}
 V_{1/2}\equiv V_{\rm circ}(r_{1/2})=\left({GM_{1/2}\over r_{1/2}}\right)^{1/2}.
\label{EqV1/2}
\end{equation}
Note that with this definition, $V_{1/2}$ is simply a rescaled measure
of the velocity dispersion, $V_{1/2}=3^{1/2}\sigma_{\rm los}$.

We note that some of the LG field galaxies and dwarf ellipticals of
M31 show some signs of rotation in their stellar component
\citep[e.g.,][]{Kirby2014,Geha2010,Leaman2012}. The implied
corrections to $M_{1/2}$ are relatively small, however, and we neglect
them here for simplicity. In addition, many of our conclusions apply
primarily to dwarf spheroidals, which are dispersion-supported systems
with no detectable rotation.

\subsection{Galaxy sample} 
\label{SecGxSample}

We use the current version of the Local Group data compilation of
\citet{McConnachie2012} as the source of our observational
dataset\footnote{More specifically, we use the October 2015 version
  from
  \url{http://www.astro.uvic.ca/~alan/Nearby\\%
    _Dwarf_Database.html}},
updated to include more recent measurements when available.  Distance
moduli, angular half-light radii, and stellar velocity dispersions
are used for estimating $V_{1/2}$ at $r_{1/2}$.  We also derive
stellar masses for all dwarfs from their distance moduli and V-band
magnitudes, using the stellar mass-to-light ratios of
\citet{Woo2008}. For cases where stellar mass-to-light ratios are not
available, we adopt $\Mstr/L_{\rm V}=1.6$ and $\Mstr/L_{\rm V}=0.7$
for dSphs and dwarf irregulars (dIrr), respectively. We list all of our
adopted observational parameters for Local Group dwarfs, as well as
the corresponding references, in Table~\ref{TabData1}.

Uncertainties in $M_{1/2}$ (or $V_{1/2}$), $\Mstr$, and $r_{1/2}$ are
derived by propagating the errors in the relevant observed
quantities. Since \citet{Woo2008} do not report individual
uncertainties on stellar mass-to-light ratios, we assume a constant 10
per cent uncertainty for all dwarfs. Our mass estimates neglect the
effects of rotation but add in quadrature an additional $20$ per cent
uncertainty to $M_{1/2}$ in order to account for the base uncertainty
introduced by the modelling procedure \citep[for details,
see][]{Campbell2016}.


Following common practice, we shall group dwarf galaxies into various loose
categories, according to their stellar mass. `Classical dSphs' is a shorthand
for systems brighter than $M_{\rm V}=-8$; fainter galaxies will be
loosely referred to as `ultra faint'. Further, we shall use the term
`faint dwarfs' to refer to all systems with $\Mstr<10^7\,
M_\odot$. The reason for this last category will become clear below.

It will also be useful to distinguish four types of
galaxies, according to where they are located in or around the Local Group:
\begin{itemize} 
\item {\bf Milky Way satellites}. These are all galaxies within 300
  kpc of the centre of the MW. Our dataset
  include all classical dSphs of the MW and all newly discovered ultra
  faint dwarfs for which relevant data are available.

\item {\bf M31 satellites}: All galaxies within 300 kpc from the
  centre of M31. Velocity dispersion measurements are available
  for many M31 satellites, mainly from \citet{Collins2013} and
  \citet{Tollerud2012}. For satellites with more than one measurement
  of $\sigma_{\rm los}$, we adopt the estimate based on the larger
  number of member stars. Structural parameters of M31 satellites in
  the PAndAS footprint \citep{McConnachie2009} have been recently
  updated by \citet{Martin2016}, whose measurements we adopt here.

\item {\bf LG field members}: These are dwarf galaxies located further
  than 300 kpc from either the MW or M31, but within 1.5 Mpc of the LG
  centre, defined as the point equidistant from the MW and M31.
 Velocity dispersion measurements are available for all of these
  systems, as reported by \citet{Kirby2014}.

\item{\bf Nearby galaxies}: These are galaxies in the compilation of
  \citet{McConnachie2012} which are further than 1.5 Mpc from the LG
  centre. This dataset includes most galaxies with accurate distance
  estimates based on high precision methods, such as the tip of the
  red giant branch (TRGB). The furthest galaxies we consider are
  located about 3 Mpc away from the MW. Velocity dispersion measurements are
  not available for all of these galaxies, but estimates exist for their
  stellar masses, half-light radii, and metallicities.
\end{itemize}

\begin{figure*}
  \hspace{-0.2cm}
  \resizebox{17.5cm}{!}{\includegraphics{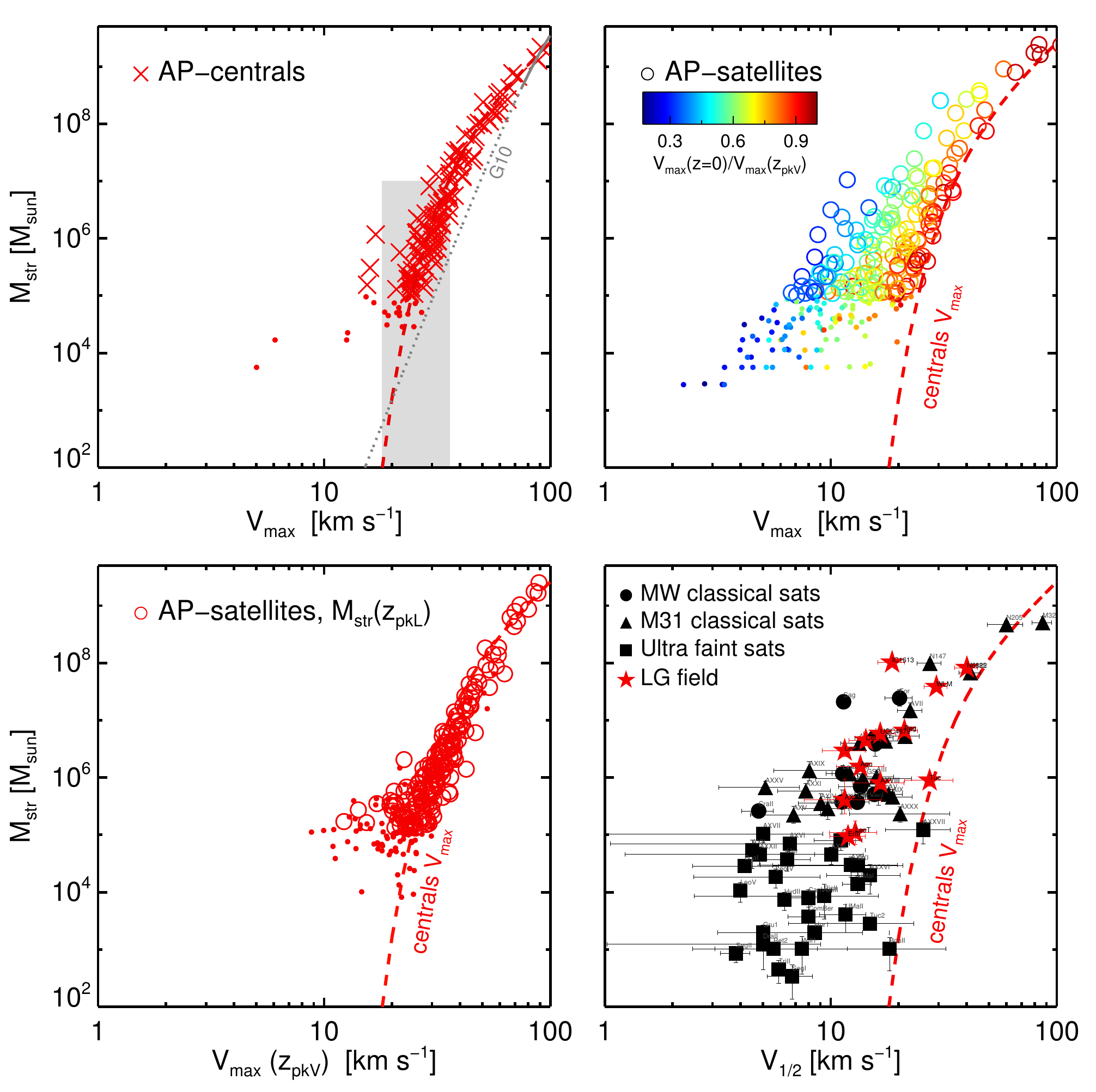}}\\%
  \caption{ {\it Top-left:} Stellar mass, $\Mstr$, versus maximum
    circular velocity, $V_{\rm max}$, of APOSTLE centrals. Crosses
    indicate all centrals $\Mstr>10^5\, M_\odot$ (resolved with more
    than $\sim 10$ particles in AP-L1 runs); dots indicate systems
    with $\Mstr<10^5\, M_\odot$ ($1$ to $10$ star particles). The dashed line is a fit of the
    form $M_{\rm str}/\Msun=m_0\,\nu^{\alpha}\exp(-\nu^{\gamma})$,
    where $\nu=V_{\rm max}/50 \kms$, and $(m_0,\alpha,\gamma)$ are
    $(3.0\times10^8, 3.36, -2.4)$. The same dashed line is repeated in
    every panel for reference. The thin grey line shows the
    extrapolation to faint objects of the abundance-matching relation
    of \citet{Guo2010}, also for reference. {\it Top-right:} Same as
    top left, but for APOSTLE {\it satellites} with
    $\Mstr>10^5\, M_\odot$. Each satellite is coloured by the reduction
    in $V_{\rm max}$ caused by tidal effects. {\it Bottom-left:}
    As top-left, but for the `peak' $\Mstr$ and $V_{\rm max}$,
    typically measured just before first accretion into the primary
    halo. {\it Bottom-right:} $\Mstr$ vs $V_{1/2}$ for LG
    dwarfs. Satellites of the MW and M31 are shown in black, `field'
    objects are shown in red. Gas-rich disc galaxies such as the
    Magellanic Clouds, M~33, or IC~10, are not considered in our
    analysis.}
\label{FigMstarVcirc}
\end{figure*}

\section{The Simulations}
\label{SecSims}

The APOSTLE project consists of a suite of zoomed-in cosmological
hydrodynamical simulations of $12$ volumes chosen to match the
main dynamical characteristics of the LG. The full selection procedure
is described in \citet{Fattahi2016} and a detailed discussion of the
main simulation characteristics is given in \citet{Sawala2016b}. 

In brief, $12$ LG candidate volumes ware selected from the DOVE dark
matter-only $\Lambda$CDM simulation of a periodic box $100 \Mpc$ on a
side \citep{Jenkins2013}. Each volume contains a relatively isolated
pair of haloes with virial\footnote{We define virial quantities as
  those contained within a sphere of mean overdensity $200\times$ the
  critical density for closure, $\rho_{\rm crit}=3H_0^2/8\pi G$, and
  identify them with a `200' subscript.} mass $M_{\rm 200}\sim10^{12}
\Msun$, separated by $d=600$--$1000$ kpc, and approaching each other
with relative radial velocity in the range $V_{\rm rad}=0$--$250
\kms$. The relative tangential velocity of the pair members was
constrained to be less than $100 \kms$, and the Hubble flow was
constrained to match the small deceleration observed for distant LG
members. Each zoomed-in volume is uncontaminated by massive boundary
particles out to $\sim 3$ Mpc from the barycentre of the MW-M31 pair.

The candidate volumes were simulated at three different levels of
resolution, labelled L1 (highest) to L3 (lowest resolution), using the
code developed for the EAGLE project \citep{Schaye2015,Crain2015}. The
code is a highly modified version of the Tree-PM/smoothed particle
hydrodynamics code, P-Gadget3 \citep{Springel2005b}. The
hydrodynamical forces are calculated using the pressure-entropy
formalism of \citet{Hopkins2013}, and the subgrid physics model was
calibrated to reproduce the stellar mass function of galaxies at
$z=0.1$ in the stellar mass range of $\Mstr=10^8-10^{12} \Msun$, and
to yield realistic galaxy sizes.

The galaxy formation subgrid model includes metallicity-dependent star
formation and cooling, metal enrichment, stellar and supernova
feedback, homogeneous X-ray/UV background radiation (hydrogen
reionization assumed at $z_{\rm reion}=11.5$), supermassive black-hole
formation, and AGN activity. Details of the subgrid models can be
found in \citet{Schaye2015,Crain2015,Schaller2015c}. The APOSTLE
simulations adopt the parameters of the `ref' EAGLE model in the
language of the aforementioned papers.

Haloes and bound (sub)structures in the simulations are found using
the FoF algorithm \citep{Davis1985} and SUBFIND \citep{Springel2001a},
respectively. First, FoF is run on the DM particles with linking
length 0.2 times the mean inter particle separation to identify the
haloes. Gas and star particles are then associated to their nearest DM
particle.  In a second step, SUBFIND searches iteratively for bound
(sub)structures in any given FoF halo using {\it all} particle types
associated to it. We shall refer to MW and M31 analogs as `primary' or
`host' haloes, even though in some of the volumes they are found
within the same FoF group. Galaxies formed in the most massive subhalo
of each distinct FoF group will be referred to as `centrals' or
`field' galaxies, hereafter.

Throughout this paper we use the highest resolution APOSTLE runs, L1,
with gas particle mass of $\sim 10^4 \Msun$ and maximum force
softening length of $134$ pc. Four simulation volumes have so far been
completed at resolution level L1, corresponding to AP-01, AP-04,
AP-06, AP-11 in table~2 of \citet{Fattahi2016}.

The simulations adopt cosmological parameters consistent with 7-year
Wilkinson Microwave Anisotropy Probe
\citep[WMAP-7,][]{Komatsu2011} measurements, as follows:
$\Omega_{\rm M}=0.272$, $\Omega_{\Lambda}=0.728$, $h=0.704$,
$\sigma_8=0.81$, $n_s=0.967$.

\begin{figure*}
  \hspace{-0.2cm}
  \resizebox{17cm}{!}{\includegraphics{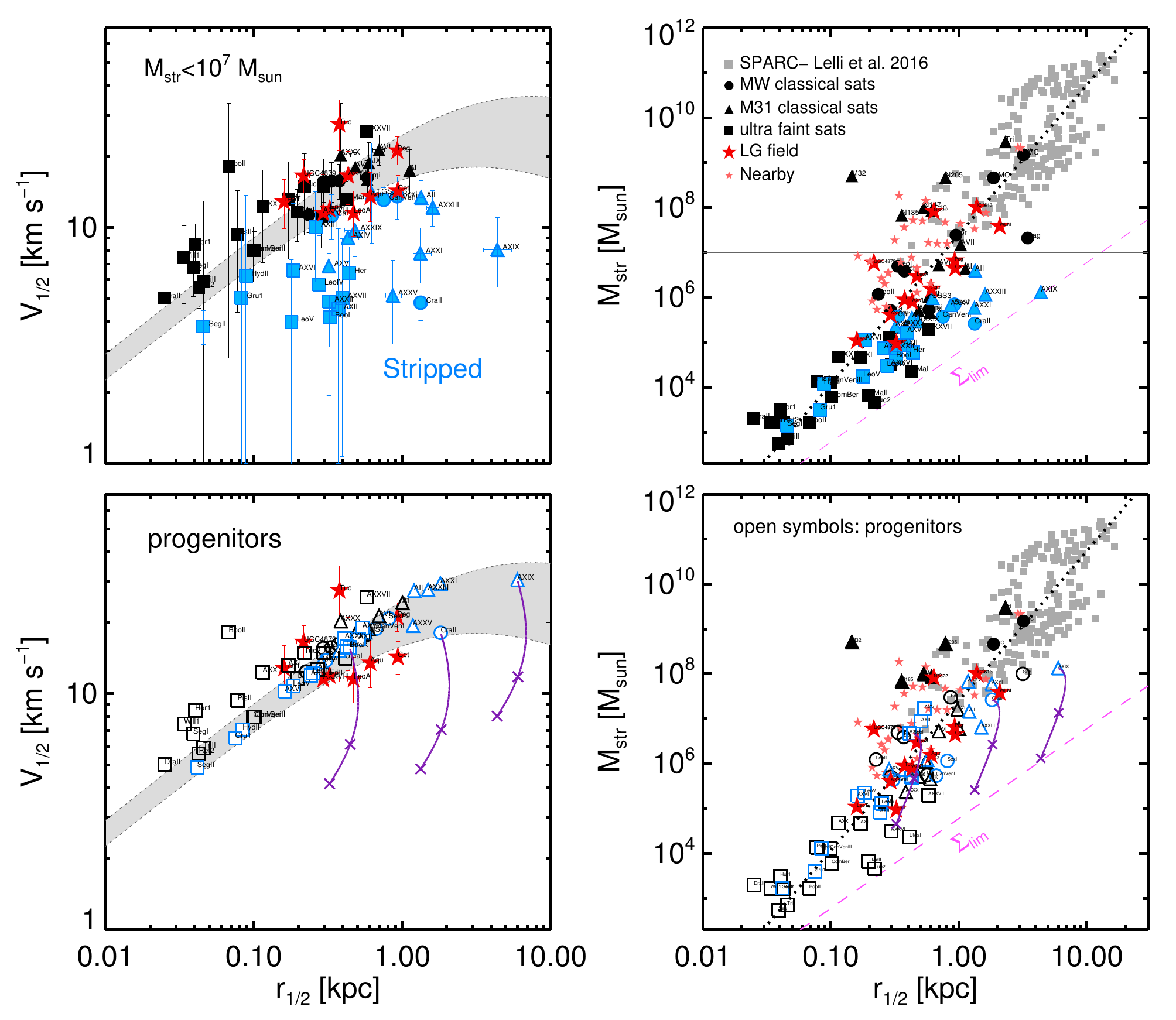}}\\%
  \caption{{\it Top-left}: Circular velocity, $V_{1/2}$, at the
    stellar half-mass radius, $r_{1/2}$, of LG `faint dwarfs'
    ($\Mstr<10^7 \Msun$), as a function of $r_{1/2}$. The shaded area
    delineates the minimum velocities expected for such dwarfs,
    bracketed by two NFW profiles with $V_{\rm max}=20$ and $36 \kms$,
    respectively (see shaded region in the top-left panel of
    Fig.~\ref{FigMstarVcirc}; symbol types are as in the bottom-right
    panel of that figure).  LG field dwarfs are shown in red, and are
    generally consistent with this expectation. Satellites with
    velocity dispersion below the shaded region are identified as
    having lost mass to tidal stripping, and are highlighted in cyan.
    {\it Bottom-left}: Same as top-left but for the progenitors of LG
    satellites, inferred as described in the text. The
    purple curves are three examples of `tidal stripping tracks'
    (PNM08). Each tick mark corresponds to successive mass losses of
    $90$ per cent. The progenitor parameters are set by assuming that
    they match $\Mstr$--$V_{\rm max}$ relation for isolated APOSTLE
    dwarfs, their $r_{1/2}$--$V_{1/2}$ follow CDM circular velocity
    profiles. (See Fig.~\ref{FigMethod} for a schematic description of
    the method.)  {\it Top-right}: $\Mstr$ vs $r_{1/2}$ relation for
    our galaxy sample as well as for the late-type galaxies in the
    SPARC survey \citep[grey squares;][]{Lelli2016a}. The dashed
    magenta line roughly indicates the minimum effective surface
    brightness limit of current surveys. {\it Bottom-right}: Same as
    top-right, but for satellite progenitors. Note that the
    progenitors are in excellent agreement with other field galaxies,
    a result that provides independent support for our proposal that
    the low-velocity dispersion satellites identified as `stripped' in
    the top-left panel have indeed been heavily affected by tidal
    stripping.}
\label{FigVcRM}
\end{figure*}

\section{Results}
\label{SecResults}

\subsection{Galaxy mass-halo mass relation in APOSTLE}
\label{SecMstarVc}

The top-left panel of Fig.~\ref{FigMstarVcirc} shows the
$\Mstr$-$V_{\rm max}$ relation for all `central' galaxies in the four
L1 APOSTLE volumes. Since we are mainly interested in dwarfs, we only
show galaxies forming in haloes with $V_{\rm max}<100 \kms$ (or,
roughly, $\Mstr<10^{10}\, M_\odot$). Galaxy stellar masses
\footnote{Stellar masses computed this way agree in general very well
  with the `bound stellar mass' returned by SUBFIND. Choosing either
  definition does not alter any of our conclusions.} are measured
within the `galactic radius', $r_{\rm gal}$, defined as
$0.15\, r_{200}$.

This panel shows the tight relation between galaxy and halo masses
anticipated for isolated APOSTLE galaxies in
Sec.~\ref{SecIntro}. Crosses indicate systems resolved with more than
$10$ star particles, and small dots systems with $1$-$10$ star
particles. It is clear that very few of the galaxies that succeed in
forming stars in our AP-L1 simulations do so in haloes with
$V_{\rm max}<20 \kms$.  In addition, essentially {\it all} isolated
`faint dwarfs' ($\Mstr<10^7\, M_\odot$) inhabit haloes spanning a
narrow range of circular velocity, $18<V_{\rm max}/\kms<36$. The few
that stray to lower velocities are actually former satellites that
have been pushed out of the virial boundaries of their primary halo by
many-body interactions \citep{Sales2007b,Ludlow2009,Knebe2011}.

The top-right panel of Fig.~\ref{FigMstarVcirc} is analogous to the
top-left, but for `satellite' galaxies\footnote{The virial radius of
  subhaloes is not well defined, so we use the average relation between
  $r_{\rm gal}$ and $\Vmax$ of centrals,
  $r_{\rm gal}/$kpc$=0.169\,(V_{\rm max}/\kms)^{1.01}$, to estimate
  the galactic radii, $r_{\rm gal}$, of satellites.}  , defined as
those within 300~kpc of either primary. The
difference with isolated systems is obvious: at fixed $\Mstr$ the
haloes of satellite galaxies can have substantially lower $V_{\rm max}$
than centrals \citep[see, also,][]{Sawala2016a}.

The difference is almost entirely due to the effect of tides
experienced by satellites as they orbit the potential of their
hosts. This is clear from the bottom-left panel of
Fig.~\ref{FigMstarVcirc}, which shows the same relation for
satellites, but for their `peak' $\Mstr$ and $\Vmax$, which
typically occur just before a satellite first crosses the virial
boundary of its host. At that time, the satellite progenitors
followed a $M_{\rm str}$-$V_{\rm max}$ relation quite similar to that of
isolated dwarfs.

Finally, the bottom-right panel of Fig.~\ref{FigMstarVcirc} shows the
stellar mass-circular velocity relation for LG dwarfs, where the
colours distinguish satellites (black) from field or isolated systems
(shown in red)\footnote{The names of Andromeda dwarfs are
  shortened in all figure legends for clarity; for example, Andromeda~XXV,
  is written as And~XXV or AXXV.}. This panel differs
from the others because the maximum circular velocity is not
accessible to observation; therefore, we show instead $V_{1/2}$, the
circular velocity at the half-mass radius (see Eq.~\ref{EqV1/2}).

The results shown in Fig.~\ref{FigMstarVcirc} elicit a couple of
comments. One is that all LG dwarfs lie to the left of the red dashed
line that delineates the $\Mstr$-$\Vmax$ relation for field APOSTLE
dwarfs. This is encouraging, since consistency with our model demands
$V_{1/2}<V_{\rm max}$ for all dark matter-dominated dwarfs.  (The only
exception is M32, a compact elliptical galaxy whose internal dynamics
are dictated largely by its stellar component.)

Second, aside from a
horizontal shift, the general mass-velocity trend of LG dwarfs is
similar to that in the simulations: below a certain stellar mass, the
characteristic velocities of LG dwarfs become essentially independent
of mass, just as for their simulated counterparts.

Finally, note that we do not show measurements of $V_{1/2}$ for
APOSTLE galaxies in Fig.~\ref{FigMstarVcirc}. This is mainly because
of the limited mass and spatial resolution of the simulations. The
majority of the LG satellites have stellar masses below $10^6 \Msun$,
which are resolved with fewer than $100$ stellar particles in even the
best APOSTLE runs, thus compromising estimates of their half-mass
radii and velocity dispersions. In addition, at very low masses, all
APOSTLE galaxies have similar, resolution-dependent, half-mass radii,
a clear artefact of limited resolution. Indeed, most AP-L1 dwarfs with
$\Mstr<10^6\, M_\odot$ have $R_{\rm eff}\sim 400$ pc
\citep{Campbell2016}. This is far in excess of the typical radii of LG
dwarfs of comparable mass, compromising direct comparisons between the
observed and simulated stellar velocity dispersions and radii of faint dwarfs.

We shall therefore adopt an indirect, but more robust, approach, where
we assume that the stellar mass-halo mass APOSTLE relation is
reliable and use it, together with the known mass profile of CDM
haloes, to interpret various observational trends in the structural
parameters of Local Group dwarfs. Our analysis thus rests on two basic
assumptions: (i) that the $\Mstr$-$V_{\rm max}$ relation of field
dwarfs follows roughly that shown in the top-left panel of
Fig.~\ref{FigMstarVcirc}; and (ii) that the baryonic assembly of the
galaxy does not alter dramatically the inner dark mass
distribution. 

The first assumption imposes a fairly sharp halo mass
  `threshold' for galaxy formation, as seen in the top-left panel of
  Fig.~\ref{FigMstarVcirc}. The existence of this threshold has been
  critically appraised by recent work, some of which argues that halos
  with masses well below the threshold 
  may form luminous galaxies \citep{Wise2014,OShea2015}, some as
  massive as the Cra~2 or Draco dwarf spheroidals \citep[see,
  e.g.,][]{Ricotti2016}. We note, however, that those simulations are
  typically stopped at high redshift ($z\sim 8$) and rarely followed
  to $z=0$, so it is unclear whether the threshold they imply (if
  expressed in present-day masses) is inconsistent with the one we
  assume here. Indeed, the latest simulation work, which includes a
  more sophisticated treatment of cooling than ours and follows
  galaxies to $z=0$, reports a comparable `threshold' to the one we use
  here \citep{Fitts2017}.

Regarding the second assumption, we emphasize that this is a
conservative one, since baryon-induced cores would only help to
reconcile CDM theoretical expectations with observations.

\begin{figure}
  \hspace{-0.3cm}
  \resizebox{8.8cm}{!}{\includegraphics{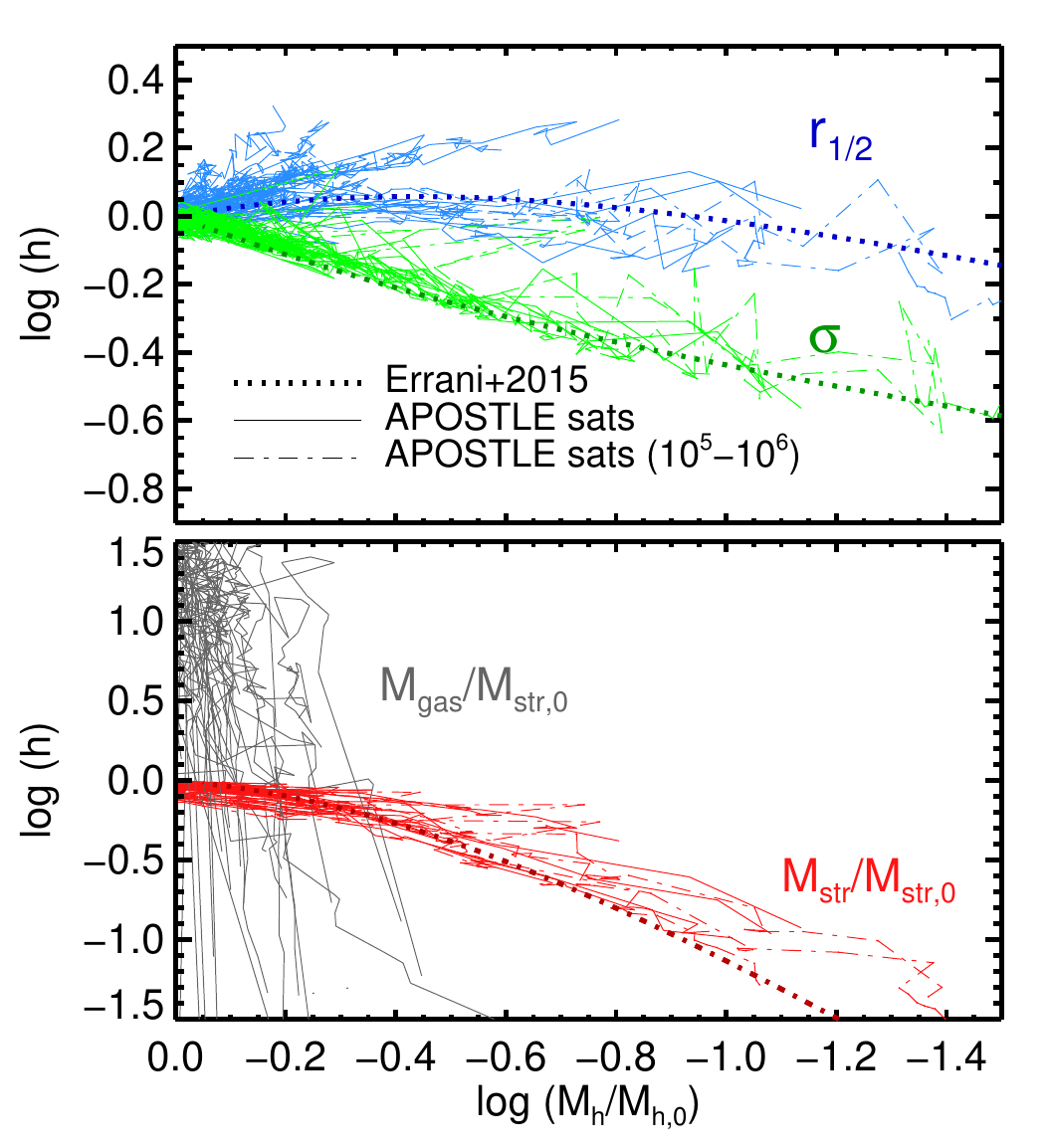}}\\%
  \caption{{\it Top}: Tidally-induced changes in the
      stellar half-mass radius ($r_{1/2}$) and stellar velocity
      dispersion ($\sigma$), as a function of the {\it total} mass
      that remains bound within the original stellar half-mass radius
      of the galaxy. The parameters are in units of their
      pre-stripping values. Thick dotted lines correspond to the
      models of \citet{Errani2015} for spheroidal galaxies embedded in
      cuspy CDM haloes. The thin solid lines indicate results for all
      APOSTLE satellites with $\Mstr>10^6 \Msun$ at present time. We
      also show, with dot-dashed lines, APOSTLE satellites with $z=0$
      stellar masses in the range $10^5-10^6 \Msun$ who have lost more
      than 90 per cent of their stellar mass in the past. 
      {\it Bottom}: Similar to the top panel but for changes in the
      stellar mass ($M_{\rm str}$) and gas mass ($M_{\rm gas}$), both
      given in units of the pre-stripping stellar mass.}
\label{FigAPstripping}
\end{figure}

\begin{figure}
  \hspace{-0.3cm}
  \resizebox{8.8cm}{!}{\includegraphics{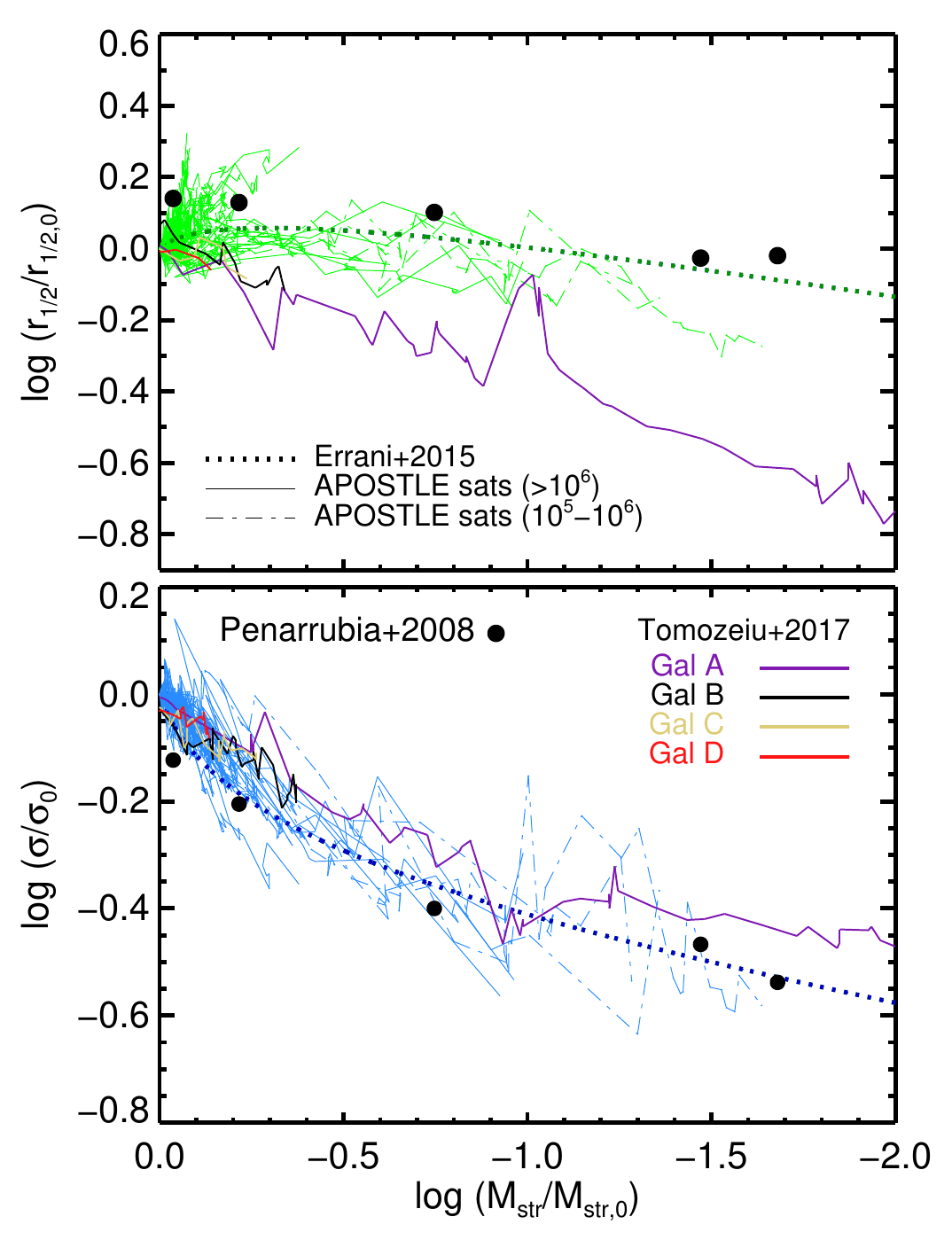}}\\%
  \caption{Tidally-induced changes in half-mass radius
      ($r_{1/2}$, top panel), and stellar velocity dispersion
      ($\sigma$, bottom panel), as a function of the remaining bound
      fraction of stellar mass. All parameters are in units of their
      pre-stripping values. Line types are as in
      Fig.~\ref{FigAPstripping}. Thick dotted curves are E15 tidal
      tracks; thin solid and dot-dashed lines are results for APOSTLE
      satellites, as in Fig.~\ref{FigAPstripping}. Solid circles
      coresspond to the six models of PNM08 at the end of their
      simulations. Thin solid lines of different colors show results
      for four disc dwarfs simulated by \citet{Tomozeiu2016}. See text
      for further discussion.}
\label{FigTomStripping}
\end{figure}

\begin{figure*}
  \hspace{-0.3cm}
  \resizebox{12cm}{!}{\includegraphics{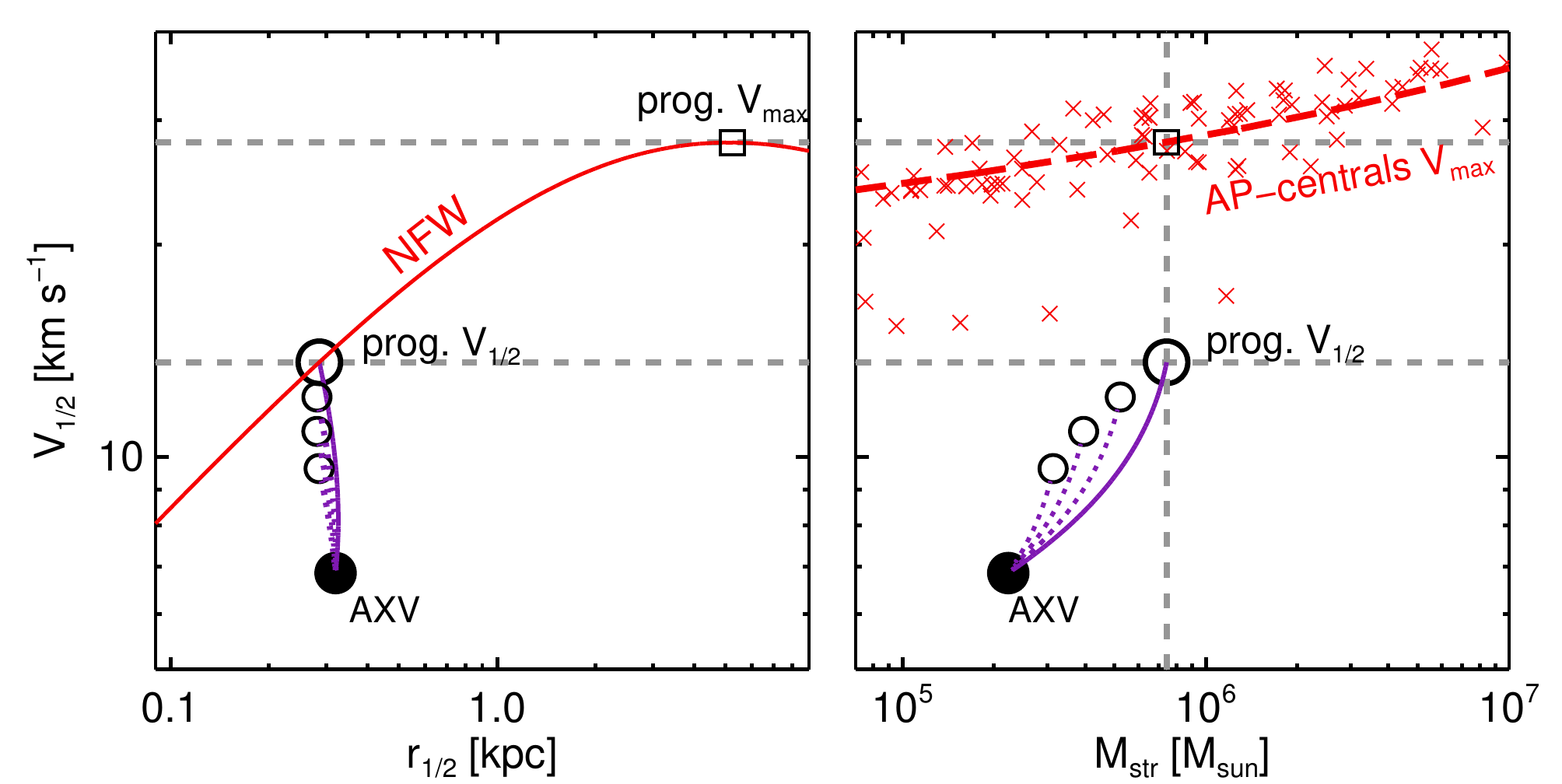}}\\%
  \caption{A schematic example to illustrate how we determine the
    properties of the progenitors of satellites deemed `stripped'
    (cyan symbols in the top-left panel of Fig.~\ref{FigVcRM}). The
    example applies to And~XV, whose present-day half-mass radius,
    circular velocity, and stellar mass are indicated by the solid
    circle. The E15 tidal tracks suggest a number of possible
    progenitors, shown by open circles. The actual And~XV progenitor
    (open square in the right-hand panel) is selected to match
    simultaneously the APOSTLE $\Mstr$-$V_{\rm max}$ relation for
    isolated dwarfs, and the circular velocity $V_{1/2}$ at $r_{1/2}$
    expected for a CDM halo of that $V_{\rm max}$ (large open circle in
    left-hand panel).}
\label{FigMethod}
\end{figure*}
  
\subsection{Tidal stripping effects on LG satellites}
\label{SecTidStrip}

\subsubsection{Size-velocity relation}
\label{SecVR}

One firm prediction of our simulations is that all dwarfs with
$\Mstr<10^7\, M_\odot$ should form in haloes of similar mass. Because
the inner circular velocity profile of CDM haloes increases with
radius, we expect the dark matter content of dwarfs to increase with
galaxy size, as larger galaxies should encompass larger amounts of
dark matter. This implies that a `minimum' velocity can be predicted
for a faint dwarf, based solely on the dark mass contained within its
half-mass radius.  This is indicated by the grey shaded region in the
top-left panel of Fig.~\ref{FigVcRM}, which indicates the dark matter
circular velocity profiles expected for haloes close to the
'threshold' (i.e., $18<V_{\rm max}/\kms <36$), modeled as NFW haloes
with concentrations taken from \citet{Ludlow2016}.

As is clear from this panel, a number of dwarfs are at odds with this
prediction, and are highlighted in cyan. Note that all of these
deviant systems are satellites (field dwarfs are shown in red). Within
the constraints of our model the only way to explain the low velocity
dispersion of these systems is to assume that they have been affected
by tides. Extreme examples include Cra~2 and And~XIX; i.e., systems
with large half-light radii and very low velocity dispersions that
are otherwise difficult to explain in our model.

\subsubsection{The progenitors of stripped satellites}
\label{SecProg}

The effects of tides on dark matter-dominated spheroidal systems
deeply embedded in NFW haloes have been explored in detail by PNM08 and
E15. One of the highlights of these studies is that structural changes
in the {\it stellar} component depend solely on the {\it total} amount
of mass lost from within the original stellar half-mass radius of a
galaxy. The fraction of stellar mass that remains bound, the decline
in its velocity dispersion, and the change in its half-mass radius are
thus all linked by a single parameter, implying that a tidally-induced
change in one of these parameters is accompanied by a predictable
change in the others.

In other words, tidally stripped galaxies trace prescribed tracks in
the space of $\Mstr$, $V_{1/2}$, and $r_{1/2}$ variables. This
restricts the parameter space that may be occupied by stripped
galaxies once the mass-size-velocity scaling relations of the
progenitors are specified.

The PNM08, or E15, `tidal tracks' may be summarized by a simple empirical
formula that describes parametrically the tidal evolution of any such
structural parameter, referred generically as $h$, in units of the original value, for a
spheroidal system deeply embedded in a cuspy (NFW) CDM halo:
\begin{equation}
h(x)={2^\alpha \, x^\beta \over (1+x)^\alpha}.
\end{equation}
Here the parameter $x$ is the {\it total} mass ($M_{\rm h}$) that
remains bound within the initial stellar half-mass radius of the
dwarf, in units of the pre-stripping value.  The values of $\alpha$
and $\beta$ are taken from E15 and given, for each
  structural parameter, in Table~\ref{TabTracks}.

\begin{table}
  \caption{Tidal evolutionary tracks according to \citet{Errani2015}}
  \bc
  \setlength{\tabcolsep}{5pt} 
  \begin {tabular*}{5.5cm}{{l} *{4}{c} }
    \hline
      &    &  $M_{\rm str}/M_{\rm str, 0}$  & $\sigma/ \sigma_0$  & $r_{1/2}/ r_{1/2, 0}$ \\
    \hline
    $\alpha$ &    & 3.57                        &  -0.68                &  1.22   \\
    $\beta$  &     & 2.06                        &  0.26                &  0.33    \\
    \hline
  \end{tabular*}
  \ec 
  \label{TabTracks}
\end{table}

We show these tidal tracks in
  Fig.~\ref{FigAPstripping} as thick dotted lines, for the case of the
  half-mass radius and velocity dispersion (top panel) and stellar
  mass (bottom).  The tracks indicate that a spheroidal galaxy that
loses $\sim 90$ per cent of its original stellar mass is expected to
experience a reduction of a factor of $\sim 2.5$ in its velocity
dispersion. On the other hand, its half-mass radius would change by
less than $20\%$. To first order, then, even if tides are able to
reduce substantially $\Mstr$ and $\sigma$, they are expected to have
little effect on the size of an NFW-embedded dwarf spheroidal.

The thin lines in Fig.~\ref{FigAPstripping} show that
  the same tidal tracks describe rather well the the change in
  $r_{1/2}$, $M_{\rm str}$, and $\sigma$ of APOSTLE satellites since
  they first cross the virial radius of their host halo. The E15 or
  PNM08 models do not include star formation, so we only consider in
  the comparison star particles born before infall. 
  We show all APOSTLE satellites with $z=0$ stellar masses exceeding
  $10^6 \Msun$ (these satellites are resolved with at least $1000$
  star particles at $z=0$), as well as those with stellar masses
  in the range $10^5-10^6 \Msun$ who have lost 90 per cent of their
  stellar mass since infall.

The agreement between the E15 models and APOSTLE
  satellites shown in Fig.~\ref{FigAPstripping} is remarkable,
  especially considering that most APOSTLE dwarfs are gas-rich at
  first infall, with gas-to-star mass ratios of order 10 to 30, and
  that the tidal tracks are only meant to decribe the evolution of the
  stellar component. Indeed, the
  gas component is lost quickly after infall as a result of tides and
  ram-pressure in the host halo \citep{Arraki2014,Frings2017}, as
  shown by the thin grey lines in the bottom panel of
  Fig.~\ref{FigAPstripping}. The gas mass loss, however, has little
  influence on the evolution of the stellar component, which remains
  close to the tidal tracks. This is because baryons never dominate
  the gravitational potential of APOSTLE dwarfs; the only parameter
  that determines the tidal evolution is the change in {\it total}
  mass, which is therefore mostly dark. The results we describe below, therefore, 
  apply mainly to dark matter-dominated dwarf spheroidals, and might need
  revision when considering systems where baryons dominate, such as,
  e.g., M32, or systems where most stars are in a thin,
  rotationally-supported disc \citep[see, e.g.,][]{Tomozeiu2016}.

Since the changes in stellar mass, velocity
  dispersion, and half-mass radius depend on a single parameter, this
  implies that they can be expressed as a function of each other. This
  is shown in Fig.~\ref{FigTomStripping}, which shows the same tracks
  as in Fig.~\ref{FigAPstripping}, but expressed as a function of the
  remaining fraction of bound stars. Here the E15 tidal tracks
  corresponding to spheroidals embedded in cuspy DM haloes (thick
  dotted lines) are compared with APOSTLE results (thin lines), as
  well as with those of PNM08 (filled circles), and with those of Gal
  A-D from \citet[][see legend]{Tomozeiu2016}. The latter authors
  embed a thin exponential disc of stars, rather than a spheroid, in a
  cuspy halo.  The E15 tracks in general reproduce well the
  tidally-induced evolution of a dwarf, except perhaps for Gal A of
  \citet{Tomozeiu2016}, which deviates from the E15 radius track when
  the stellar mass loss is extreme (i.e., more than 90 per cent). We
  note, however, that the few APOSTLE dwarfs who suffer comparable
  stellar mass loss seem to agree with the E15 tracks quite well. The
  difference is likely due to the fact that the initial galaxies in
  \citet{Tomozeiu2016} are pure exponential discs rather than
  spheroids, but further simulations would be needed to confirm this.

One important corollary of these results is that the E15 tidal tracks
can be used to `undo' the effects of stripping once the structural
properties of the progenitors are specified.  We attempt this in the
bottom-left panel of Fig.~\ref{FigVcRM}, where we show the $V_{1/2}$
vs $r_{1/2}$ relation for the progenitors of all LG satellites,
assuming that they follow the APOSTLE scaling relations appropriate
for isolated dwarfs (i.e., top-left panel of
Fig.~\ref{FigMstarVcirc}).

A detailed, schematic example of the procedure is presented in
Fig.~\ref{FigMethod} for the case of And~XV: the properties of the
progenitor are uniquely specified once it is constrained to match
simultaneously the $\Mstr$--$V_{\rm max}$ relation expected of APOSTLE
isolated dwarfs and the $r_{1/2}$--$V_{1/2}$ relation, assuming NFW
mass profiles. `Progenitors' computed this way will be shown with open
symbols in subsequent figures\footnote{We do not track
  baryon-dominated satellites, M32, NGC 205, NGC 147, and NGC 185,
  since our procedure applies only to dark matter-dominated
  systems. For the Sagittarius dSph we assume that the progenitor has
  a luminosity of $10^8 \Msun$, following
  \citet{Niederste-Ostholt2010}.}. The parameters of LG satellites and
their assumed progenitors are listed in Tables~\ref{TabData2} and
\ref{TabProg}.

The tracks in the bottom-left panel of Fig.~\ref{FigVcRM} highlight
three systems which, according to our procedure, have been very heavily
stripped: Cra~2, And~XIX, and Boo~I. A tickmark along each track
indicates successive factors of $10$ in stellar mass loss. For most
satellites the procedure suggests modest mass losses, but for these
three (rather extreme) examples our procedure suggests that each has
lost roughly $99$ per cent of their original mass.

\subsubsection{Mass-size relation}
\label{SecMR}

The discussion above suggests that tides have had non-negligible
effects on many LG satellites.  Is there any independent supporting
evidence for this conclusion? One possibility is to examine how other
scaling laws are affected by the changes in velocity and radius
prescribed by our progenitor-finding procedure. We emphasize that this
procedure is based on a {\it single} assumption (aside from assuming
NFW mass profiles for the progenitors): that all satellites descend
from progenitors that follow the $\Mstr$-$\Vmax$ relation for isolated
dwarfs in APOSTLE.

We begin by examining, in the top-right panel of Fig.~\ref{FigVcRM},
the stellar mass versus half-light radius relation for our whole
galaxy sample, enlarged by the late-type galaxies from the SPARC
sample\footnote{Following \citet{Lelli2016a}, we assume a stellar
  mass-to-light ratio of 0.5 in the 3.6 $\mu m$ band for SPARC
  galaxies.} of \citet{Lelli2016a}. Galaxy size and mass are clearly
correlated ($M\propto r^{2/7}$; thick dotted line), so that the
effective surface brightness increases roughly as
$\Sigma \propto M^{3/7}$.  There is also substantial scatter in radii
at fixed stellar mass, and vice versa.

An interesting feature of this plot is the clear separation between
the satellites deemed `stripped' because of their low velocity
dispersion (shown in cyan) and field LG dwarfs (shown in
red). Although there is little overlap in stellar mass, satellites and
field LG dwarfs do overlap in size. Satellites, however, appear to
follow a different trend in the mass-radius plane than that of the
general population (shown with a dashed line in the top-right panel of
Fig.~\ref{FigVcRM}. In our interpretation, this {\it difference in
  mass at fixed radius} is a signature of tidal stripping, and
should disappear when considering the properties of their progenitors.

We show this in the bottom-right panel of Fig.~\ref{FigVcRM}, where
we can see that the mass and size of the {\it progenitors} are in excellent
agreement with the general population of field galaxies. In other
words, the same correction in velocity dispersion required to
restore agreement with APOSTLE predictions for isolated dwarfs also
brings the population of `stripped' satellites into agreement with
the general field population in terms of stellar mass and size. We emphasize
that there is no extra freedom in this procedure. Once the change in
velocity dispersion is specified, the change in radius and mass
follows, as illustrated by the stripping tracks in Fig.~\ref{FigAPstripping}.

This exercise offers a simple explanation for why satellites as faint
and kinematically cold as Cra~2 and And~XIX are so large in size: they
are the tidal descendants of once more massive systems, which were
born physically large and have remained so even after being heavily
stripped. Recall that, according to the stripping tracks of PNM08 and
E15, the size of the stellar component of a dSph embedded in an NFW
halo is affected little by stripping, even after losing $\sim 99$ per
cent of its original stellar mass.

Note as well that {\it not all satellites are strongly stripped}, and that
those that have been stripped have been affected to varying
degrees. This is not unexpected, since the effectiveness of stripping
depends sensitively on the mass of the satellite; on how concentrated
the stellar component is within its halo; on the pericentric distance
of its orbit; and on the number of orbits it has completed. All of
those parameters can vary widely from system to system, scrambling the
original $r_{1/2}$-$V_{1/2}$ correlation (bottom-left panel of
Fig.~\ref{FigVcRM}) and turning into the largely 
scatter plot we see in the top-left panel of the same figure.

\begin{figure}
  \hspace{-0.25cm}
  \resizebox{8.8cm}{!}{\includegraphics{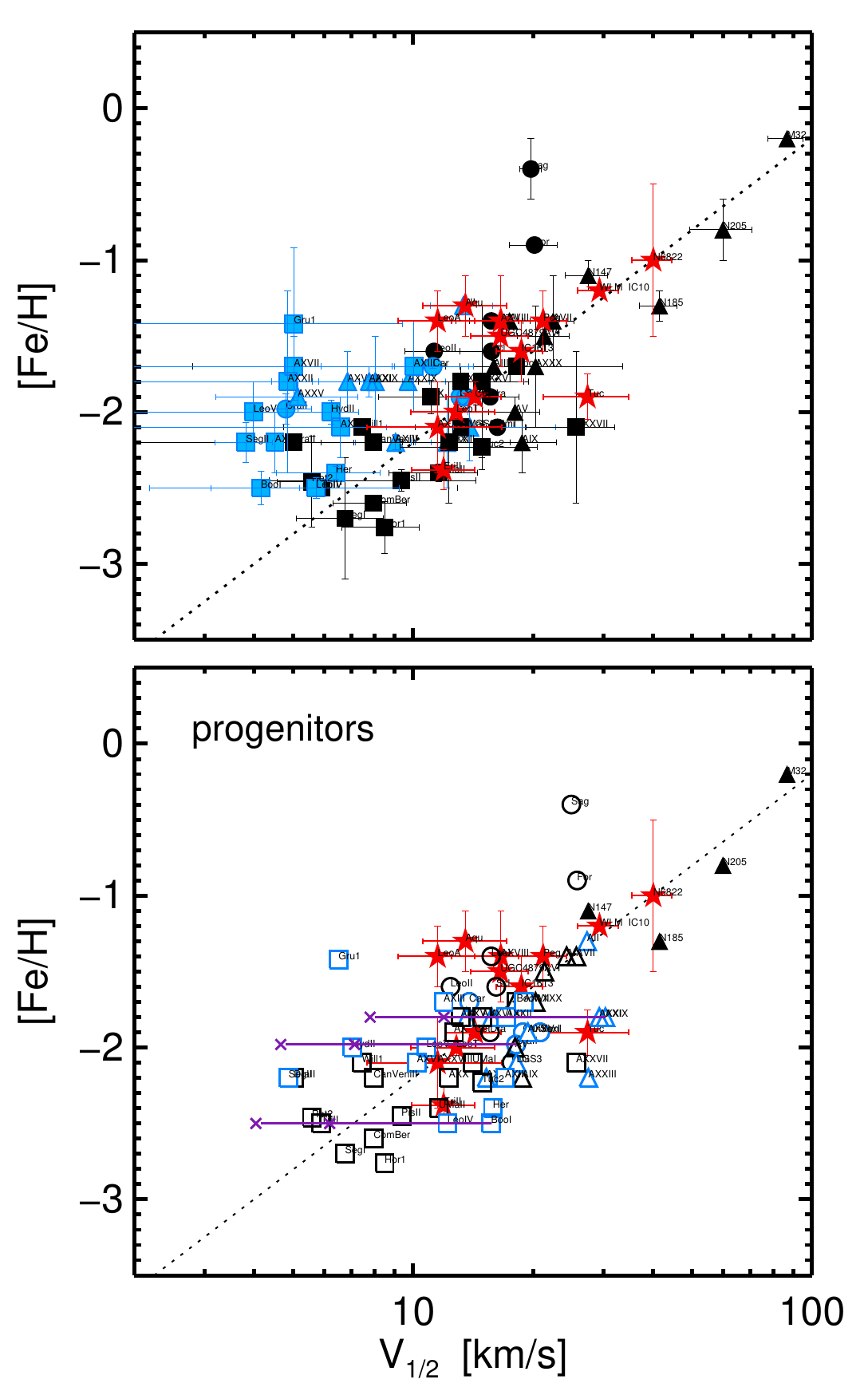}}\\%
  \caption{{\it Top}: [Fe/H] vs $V_{1/2}$ for dwarf galaxies in the
    LG. Symbol types and colours are as in Fig.~\ref{FigVcRM}. The
    stripped satellites (cyan symbols), contribute a population that
    flattens the relation at the low-velocity end. Satellites deemed
    `stripped' have lower velocity dispersions than field dwarfs (red
    symbols) of comparable metallicity.  {\it Bottom}: As top panel,
    but for satellite progenitors, assuming that their metallicities
    are unaffected by tides (i.e., they shift only horizontally in
    this plot). The tidal stripping correction restores agreement
    between satellites and field galaxies, and result in a tighter,
    monotonic relation between metallicity and velocity for all
    dwarfs.}
\label{FigFeV}
\end{figure}

\subsubsection{Metallicity-velocity dispersion relation}
\label{SecFeSigma}

Tidal stripping is expected to affect the least scaling
laws involving the metallicity of a dwarf, which would only be
modified in the case of a pronounced metallicity gradient in the
progenitor. Assuming, for simplicity, that tidal losses leave the
average metallicity of a satellite unchanged, we examine the effects of
stripping on the relation between metallicity and velocity
dispersion. We prefer to use velocity dispersion instead of stellar
mass because, according to the tidal tracks of E15 or PNM08, changes in
velocity are a more sensitive measure of tidal stripping than changes
in stellar mass.

This is shown in the top panel of Fig.~\ref{FigFeV} for all galaxies in
our sample (Sec.~\ref{SecGxSample}) with published
measurements of these two quantities.  We use in this panel the latest
observed metallicities, but caution that some are estimated
spectroscopically from individual stars whereas others rely on
photometric estimates based on the color of the red giant branch
\citep[see ][and references therein]{McConnachie2012}. 
There is a reasonably well defined trend of
increasing metallicity, [Fe/H], with increasing $V_{1/2}$, except at
the low velocity end, where the trend falters and the relation turns
flat.

The flattening is largely a result of the low-velocity population that
we have identified as `stripped' satellites (shown in blue in
Fig.~\ref{FigFeV}). Interestingly, the trend between velocity and
metallicity for progenitors is monotonic and tighter when considering
their inferred progenitors (bottom panel of the same figure), lending
further support to our assumption that the low-$V_{1/2}$ population
originates from tides.

\begin{figure}
  \hspace{-0.2cm}
  \resizebox{8.8cm}{!}{\includegraphics{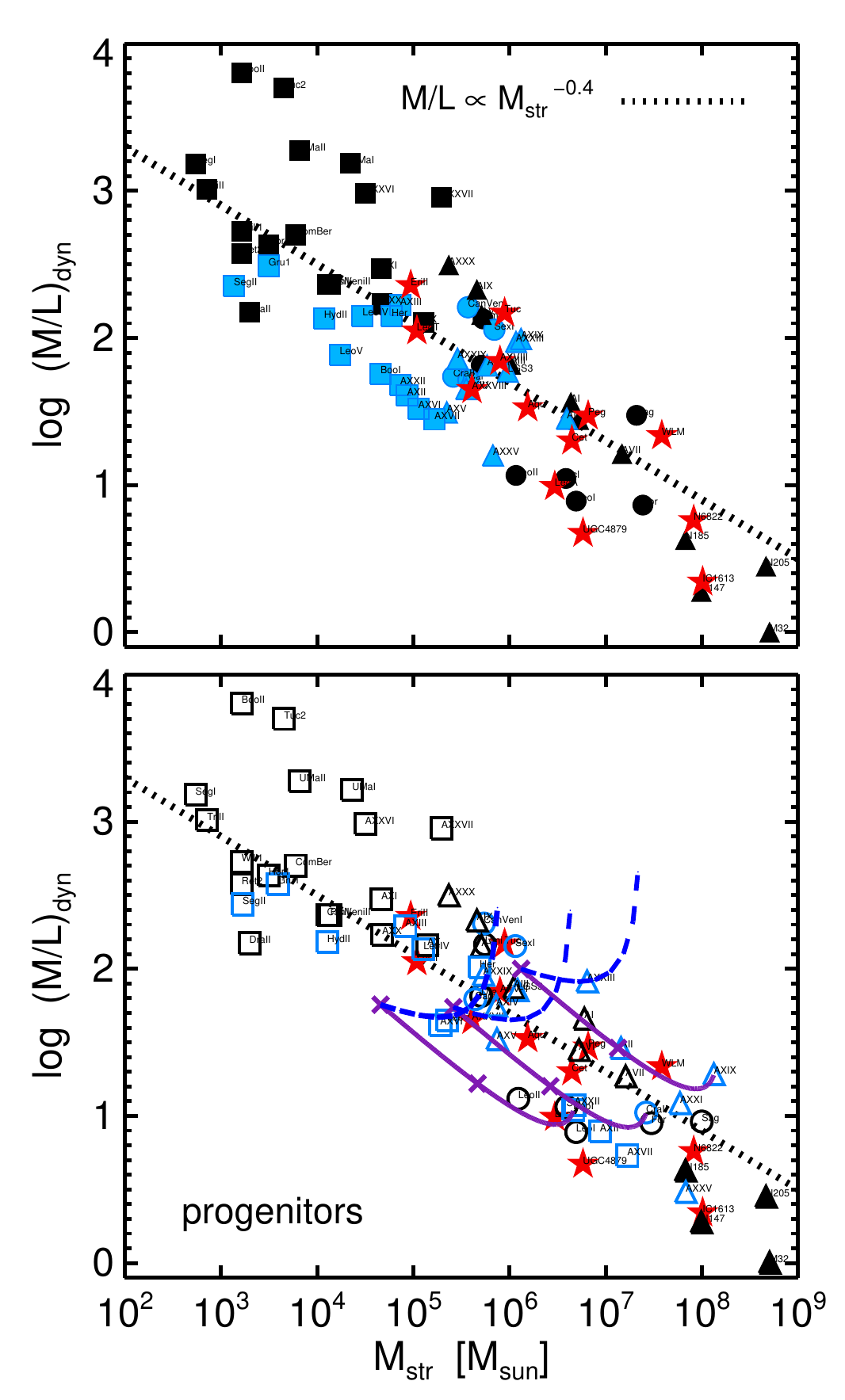}}\\%
  \caption{As Fig.~\ref{FigFeV}, but for the stellar mass vs dynamical
    mass-to-light ratio relation. The top panel shows the results for
    LG dwarfs; bottom panel for their inferred progenitors. Note that
    tidal stripping moves satellites along tracks parallel to the
    observed relation, so that stripped and unstripped systems follow
    the same relation. The thick dotted lines show $(M/L)_{\rm dyn}
    \propto M^{-0.4}$, motivated by the $V\propto r^{1/2}$ relation
    expected for the inner regions of an NFW halo, together with the
    $L\propto r^{7/2}$ scaling that holds for field galaxies (see
    top-right panel of Fig.~\ref{FigVcRM}). The blue dashed lines
    represent tidal tracks for a model in which the dark matter halo
    has a central core of size comparable to the size of the
    corresponding stellar component.}
\label{FigMLratio}
\end{figure}

\subsubsection{Dynamical mass-to-light ratios}

One firmly established dwarf galaxy scaling law links the
dynamical mass-to-light ratio, $(M/L)_{\rm dyn}\equiv M_{1/2}/(L_V/2)$, with
the total luminosity. As discussed in the early review by
\citet{Mateo1998}, dSphs have mass-to-light ratios that increase
markedly with decreasing luminosity, `consistent with the idea that
each is embedded in a dark halo of fixed mass'. How is this relation
modified by our proposal that tidal stripping may have 
altered the size, stellar mass, and velocity dispersion of many
satellites?

We examine this in Fig.~\ref{FigMLratio}, where the top panel shows
the dynamical mass-to-light ratios of all LG galaxies in our sample,
as a function of stellar mass. 
Interestingly, tidal stripping does not alter this overall scaling, as
it mainly shifts galaxies along lines roughly parallel to the main
trend. Indeed, the progenitors sample a very similar relation as the
present-day satellites, as may be seen in the bottom panel of
Fig.~\ref{FigMLratio}. As discussed by PNM08, this is a result of the
particular tidal stripping tracks expected for stellar
systems embedded in `cuspy' NFW haloes. 

If dark matter haloes had instead constant density cores comparable in
size to the stellar component, then the change in mass-to-light ratio
due to tidal stripping for a given change in stellar mass would be
much more pronounced. This is shown by the blue dashed lines, which
indicate the tidal tracks expected in such a case, as given by E15.
Had some satellites lost a large fraction of their original mass to
tides, they would have moved away from the $(M/L)_{\rm dyn}$-$\Mstr$
relation that holds for the progenitors. On the other hand, if haloes
are `cuspy' then tidally-stripped galaxies just move along the
observed relation: isolated dwarfs, progenitors, and tidal remnants
are all expected to follow the same relation.

\begin{figure}
  \hspace{-0.2cm}
  \resizebox{8.8cm}{!}{\includegraphics{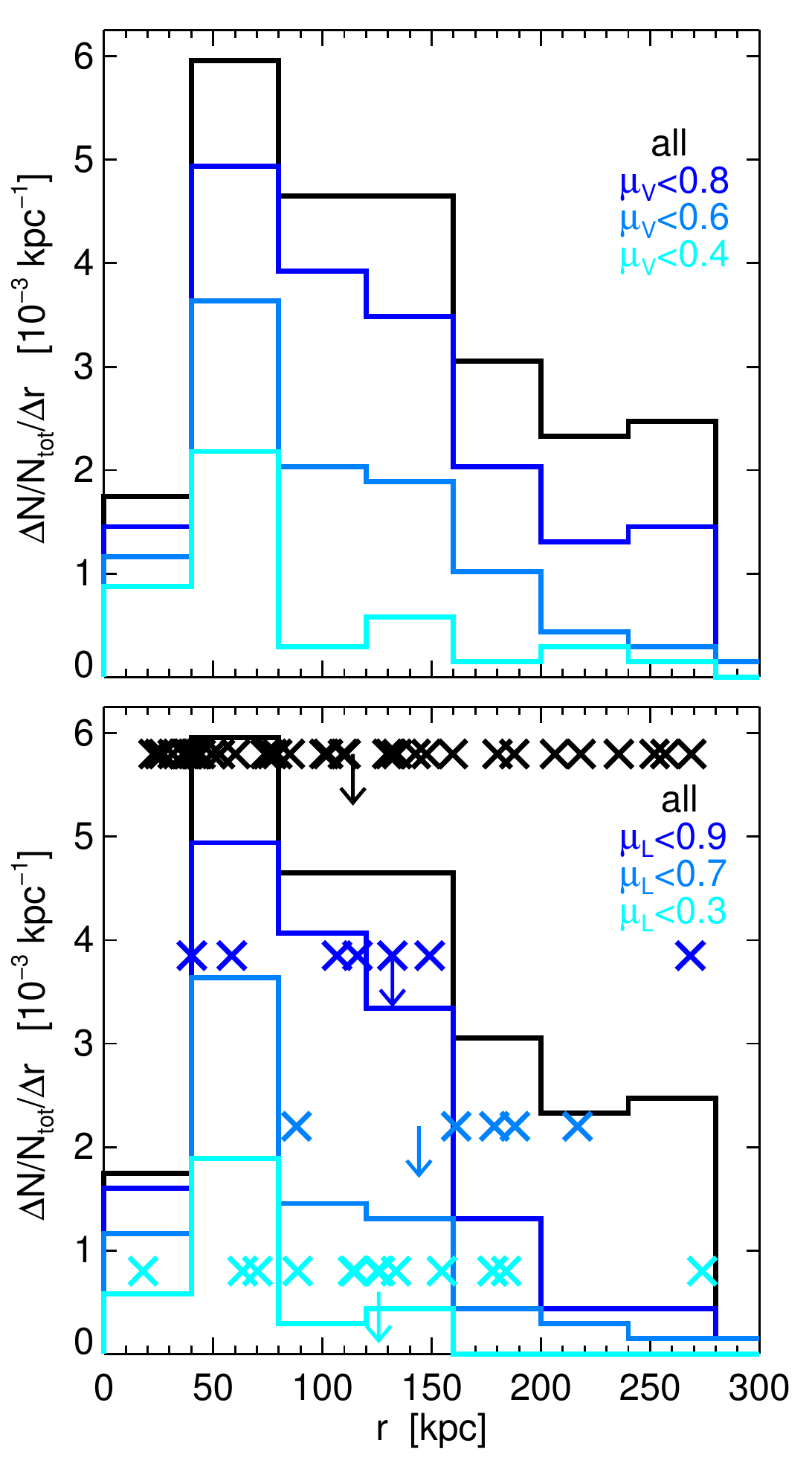}}\\%
  \caption{{\it Top}: Radial distribution of all APOSTLE satellites
    with $\Mstr>10^5\, M_\odot$ (black curves). Lower coloured histograms
    correspond to `stripped' systems, as estimated by the parameter
    $\mu_{\rm v}$, which measures the decline in $V_{\rm max}$ caused
    by tides (see text for details). {\it Bottom:} Same as top, but
    for the stripping parameter $\mu_{\rm L}$, which measures the loss
    in stellar mass caused by tides. Note that highy stripped systems
    are more centrally concentrated than the average satellite
    population. Crosses indicate the location of LG satellites,
    coloured by their inferred tidal mass loss, as described in
    Sec.~\ref{SecProg}, and summarized in Table~\ref{TabProg}. See
    text for further discussion.}
\label{FigRdist}
\end{figure}

\subsubsection{Tidal stripping and satellite shapes}

Our discussion above suggests that the observed dwarf galaxy scaling
laws pose no fundamental problem to a scenario where tides have
affected a number of satellites, even if in some cases, such as
Cra~2 and And~XIX, the posited fraction of mass lost may approach
$99$ per cent. Two oft-cited arguments against this scenario involve
satellite shapes and their distances to the primary galaxy. 

Cra~2, for example, is rather round on the sky, and it is today
situated at $\sim 115$ kpc from the Galactic Centre
\citep{Torrealba2016}. Do such observations contradict our idea that
Cra~2 has lost many of its original stars to tides?

Not necessarily. First, we should recall that the idea that heavily
stripped systems must be very aspherical only applies to systems near
the pericentre of their orbits and thus `caught in the act' of being
stripped, such as, for example, the Sagittarius dSph
\citep{Ibata2001,Majewski2003}, and the globular cluster Pal~5
\citep{Odenkirchen2001,Odenkirchen2003}. These are clearly convincing
examples of the effect of Galactic tides, but not typical.

Indeed, we expect most satellites to be on rather eccentric orbits
around the Galactic Centre, which means that tidal effects are best
approximated as impulsive perturbations that operate at pericentre. As
discussed by \citet{Penarrubia2009}, the signature of Galactic tides
fades away from the bound remnant quickly (i.e., within one crossing
time) after pericentric passage.  This implies that the effect of
tides is actually rather difficult to discern when the satellite is at
apocentre, where it spends most of its orbital time and is therefore
most likely to be found.

In addition, tidal remnants are expected to be {\it much rounder} than
their progenitors when equilibrium has been restored \citep[see;
e.g.,][and references therein]{Barber2015}. Tides actually tend to
reduce the original asphericity of a galaxy, implying that there is in
principle no contradiction between round satellite shapes and the
possibility of heavy tidal stripping.

\subsubsection{Tidal stripping and satellite spatial distribution} 
\label{SecRdist}

Satellites that have been extremely affected by
tides are expected to be in orbits with small pericentric distances
and should have completed at least a few orbits around the primary
galaxy. The latter condition implies either a small apocentre or an
early time of accretion into the primary halo, or both. One may
therefore argue that the large distances from the Galactic Centre of
some low velocity dispersion satellites are inconsistent with a tidal
origin for their peculiar properties.

We examine this in APOSTLE, where we can easily identify systems that
have experienced substantial tidal mass loss, track their orbits, and
compute their orbital parameters. We explore two alternative measures
of tidal stripping for subhaloes that, at $z=0$, still host a luminous
satellite: one is the reduction in $V_{\rm max}$ experienced since
accretion; the other is the {\it stellar} mass loss since the peak of
stellar mass.

Neither measure is ideal. The first one suffers from the fact that
$V_{\rm max}$ changes are sensitive mostly to the tidal loss of dark
matter, which couples in a complex and indirect way to actual stellar
mass losses. The second quantity measures directly stellar mass losses
but is vulnerable to numerical artefact, since the mass loss is
expected to depend sensitively on the stellar half-mass radii, which
are poorly resolved in APOSTLE, especially at the faint end (see
discussion in Sec.~\ref{SecMstarVc}).

We therefore pursue both alternatives in our analysis, and show the
results in Fig.~\ref{FigRdist}.  Because of the caveats above, this is
only meant to identify possible major inconsistencies in our argument, rather
than to provide quantitative estimates that can be directly compared
with observations.

The top panel of Fig.~\ref{FigRdist} shows, in black, the radial
distribution of all 
$\Mstr>10^{5}\, M_\odot$ satellites found, at $z=0$,
within $300$ kpc from the centre of AP-L1 primaries. The
luminous satellite radial distribution is also shown for several
subsamples, drawn according to the tidally-induced reduction of the maximum
circular velocity of each subhalo, measured by the ratio
$\mu_{\rm v}=V_{\rm max}(z=0)/V_{\rm max}(z_{\rm pkV})$. Here
$z_{\rm pkV}$ identifies the time when $V_{\rm max}$ peaked, which
typically occurs just before being first accreted into the primary
halo. 

The various distributions in the top panel of Fig.~\ref{FigRdist}
(labelled by $\mu_{\rm v}$) show the radial segregation of satellites
that have been heavily affected by tides. Clearly, the larger the
effects of tides, the closer to the galaxy centre satellites lie, on
average. Note that heavily stripped systems are not particularly rare:
$18$ per cent of all subhaloes with satellites as massive as
$\Mstr>10^{5}\, M_\odot$ have $\mu_{\rm v}<0.4$. This corresponds to a
rather large ($>95$ per cent) loss of the original total bound mass
(see PNM08's Fig.~8). Note that some of these very highly stripped
objects may be found quite far from the centre of the primary, even as
far out as $\sim250$~kpc.

The bottom panel of Fig.~\ref{FigRdist} is analogous to that in
the top, but adopting the ratio
$\mu_{\rm L}=\Mstr (z=0)/\Mstr (z_{\rm pkL})$. Here $z_{\rm pkL}$
identifies the time when the {\it stellar} mass of a satellite
peaked. The various distributions, labelled by the corresponding
values of $\mu_{\rm L}$, show that heavily stripped systems are not
particularly rare. Of all surviving luminous satellites in APOSTLE,
more than $13$ per cent have lost $>70$ per cent of their stars (i.e.,
$\mu_{\rm L}<0.3$), but we caution again that this number is rather
uncertain because of limited resolution.  The sequence of histograms
in the right panel of Fig.~\ref{FigRdist} again shows that 
highly stripped satellites tend to be more centrally concentrated than
the average.

We compare this with our stripping estimates for the LG satellite
population by indicating with crosses the distance to the primary
(MW or M31) of all satellites (in black) and of those deemed,
according to our progenitor-finding procedure, to have lost various
fractions of their original mass (in colour; each satellite is only
plotted once, and the median of each population is shown with a small
arrow).

Focusing on the most highly-stripped population (i.e., $\mu_{\rm
  L}<0.3$) we note that most of them are well within $150$~kpc of the
centre, both in the observations and in the simulations. We conclude that
there is no obvious inconsistency between the spatial distribution of
low-velocity dispersion satellites and our hypothesis that their
peculiar properties have been caused by tidal stripping.

\begin{figure*}
  \hspace{-0.2cm}
  \resizebox{17cm}{!}{\includegraphics{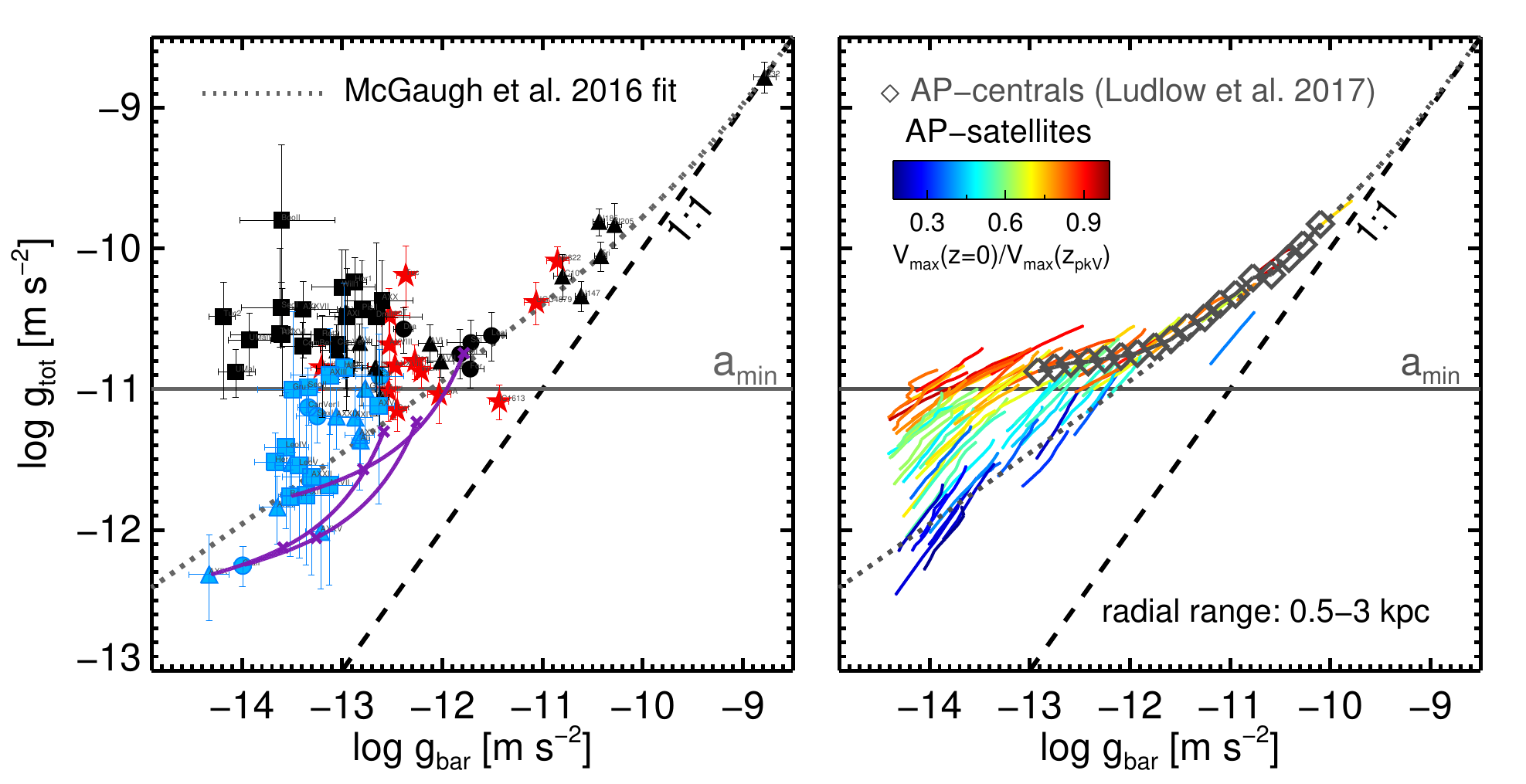}}\\%
  \caption{{\it Left}: The acceleration, $g_{\rm
      tot}=V_{1/2}^2/r_{1/2}$, at the stellar half-mass radius, as a
    function of the baryonic contribution at that radius, $g_{\rm
      bar}=G\Mstr/2\,r_{1/2}^2$, computed assuming spherical
    symmetry. The symbols show results for all LG dwarfs, using the
    same colours and types as in Fig.~\ref{FigVcRM}. The thick
    dotted line is the empirical MDAR fit of \citet{McGaugh2016}, as
    given by Eq.~\ref{EqGtotGbar}. The horizontal line highlights
    $a_{\rm min}$, the minimum acceleration expected for isolated
    dwarfs in $\Lambda$CDM \citep{Navarro2016}. Tidal stripping is
    expected to push some satellites below that minimum, as shown by
    the tidal tracks shown in magenta. Note the large scatter at the
    low-$g_{\rm bar}$ end. {\it Right:} As left panel but for the
    average of all APOSTLE central (`field') galaxies \citep[connected
      squares, as given by][]{Ludlow2017}. Coloured red lines
    illustrate the expected location of APOSTLE satellites in this
    panel. Since the stellar half-mass radii of faint simulated
    satellites is poorly constrained, we show for each subhalo a line
    segment that spans a wide range in radius, $0.5<r/{\rm kpc}<3$,
    covering the full observed range in $r_{1/2}$ at given
    $\Mstr$. Each subhalo is coloured by the tidal stripping measure
    $\mu_{\rm v}$ introduced in Sec.~\ref{SecRdist}, which measures
    the decline in $V_{\rm max}$ caused by stripping. Note that
    satellites are expected to `fan out' at low values of $g_{\rm
      bar}$, as observed in the left-hand panel.}
\label{FigMDAR}
\end{figure*}

\begin{figure*}
  \hspace{-0.2cm}
  \resizebox{17cm}{!}{\includegraphics{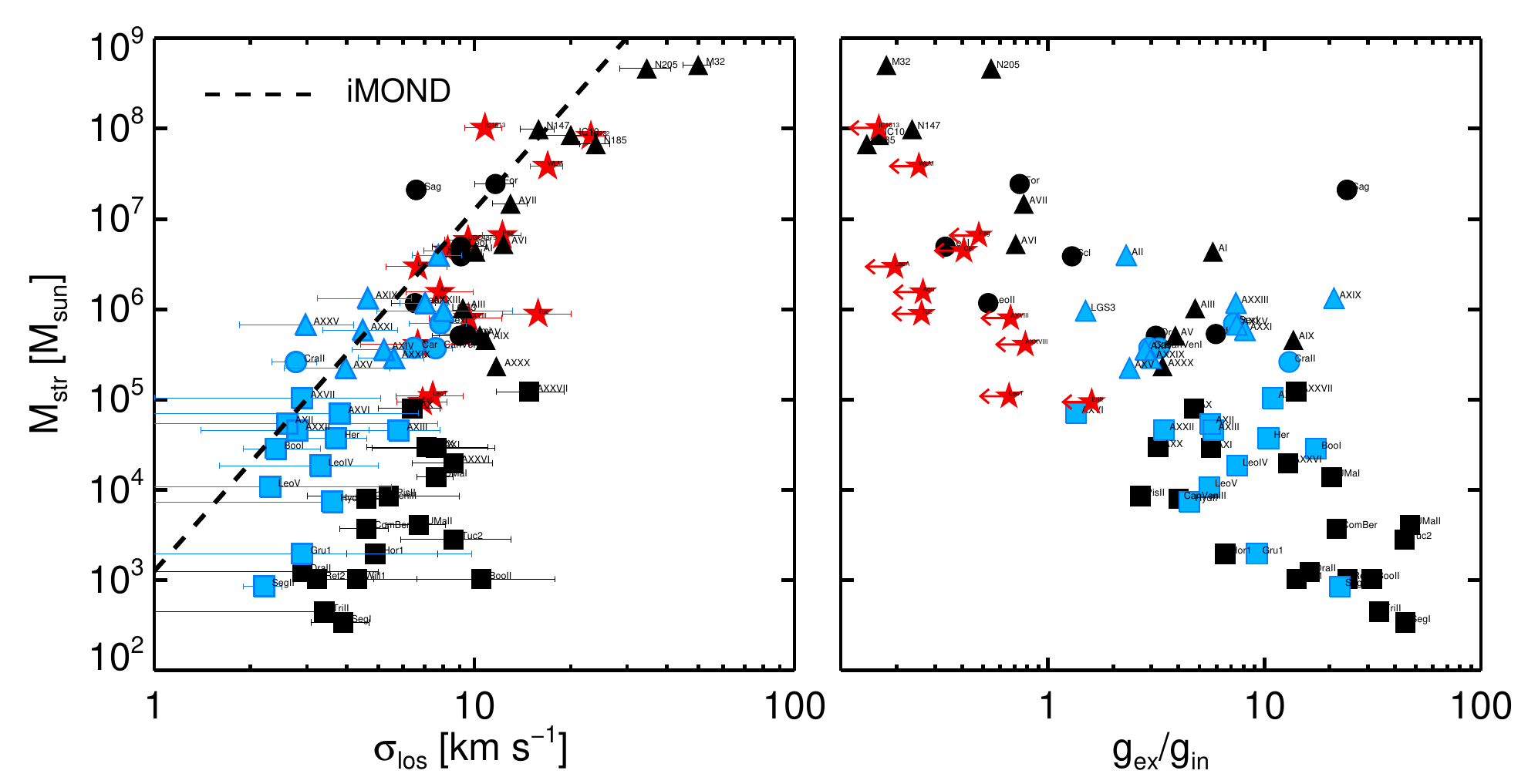}}\\%
  \caption{{\it Left}: Stellar mass-velocity dispersion relation for
    all LG dwarfs. Symbols and and colours are as in
    Fig.~\ref{FigVcRM}. The thick dotted line is the MOND prediction
    for isolated systems, as in Eq.~\ref{EqSigmaiMond}. Note that many
    faint galaxies have velocity dispersions well in excess of what is
    predicted by MOND. {\it Right:} Stellar mass as a function of the
    ratio of `external' to `internal' accelerations,
    $g_{\rm ex}/g_{\rm in}$. This provides a measure of the importance
    of `external field effects' (EFE) on MOND predictions.}
\label{FigMOND}
\end{figure*}

\subsection{Tidal stripping and the MDAR}
\label{SecMDAR}

One consequence of the effects of tidal stripping discussed in the
previous subsection is that stripping is expected to scatter satellite
galaxies away from the `mass discrepancy-acceleration relation' (MDAR)
that holds for isolated galaxies. Various forms of this relation
have been proposed in the past, but we adopt for our discussion here
the latest results of \citet{McGaugh2016} and \citet{Lelli2016a}.

These authors show a tight correlation between the gravitational
acceleration estimated from the rotation curve of late-type galaxies,
$g_{\rm tot}=V_{\rm rot}^2(r)/r$, and the acceleration expected from
the luminous (baryonic) component of a galaxy,
$g_{\rm bar}=V_{\rm bar}^2(r)/r$, where $V_{\rm bar}(r)$ is the
contribution of the baryons to the circular velocity at radius
$r$. The relation may be approximated by the fitting function,
\begin{equation}
g_{\rm tot}={g_{\rm bar} \over 1-e^{-\sqrt{g_{\rm bar}/g_\tau}}}\,,
\label{EqGtotGbar}
\end{equation}
over the range $-11.7 < \log (g_{\rm bar}/{\rm m \, s}^{-2})<-9$,
with relatively small residuals.

At the (faint) low-$g_{\rm bar}$ end\footnote{Note that $g_{\rm bar}$
  is roughly proportional to the surface brightness of a galaxy. Since
  surface brightness generally decreases with decreasing luminosity,
  faint dwarfs typically populate the low-$g_{\rm bar}$ end of the
  relation.}, the relation seems to flatten, with $g_{\rm tot}$
approaching an asymptotic minimum value of
$a_{\rm min}\sim 10^{-11} {\rm m\, s}^{-2}$ \citep{Lelli2016}. This
flattening has been called into question by the Cra~2 dSph, which
seems to lie on the extrapolation of Eq.~\ref{EqGtotGbar} \citep{McGaugh2016b}, at
$(g_{\rm bar},g_{\rm tot})=(1.0\times 10^{-14}, 5.6\times 10^{-13})$, with
all accelerations measured in ${\rm m\, s}^{-2}$.

This issue is of interest to our discussion, since $\Lambda$CDM dwarf
galaxy formation models such as that of APOSTLE make a very specific
prediction for this relation: the minimum halo mass threshold
discussed in Sec.~\ref{SecMstarVc} to host a luminous dwarf translates
into a well-defined minimum acceleration that all {\it isolated}
dwarfs must satisfy. As discussed in detail by \citet{Navarro2016} and
\citet{Ludlow2017}, this minimum acceleration is of order $a_{\rm
  min}\sim 10^{-11} {\rm \, m\, s}^{-2}$, which provides a natural and
compelling explanation for the faint-end flattening of the relation
reported by \citet{Lelli2016}.

We illustrate the simulation predictions in the right-hand panel of
Fig.~\ref{FigMDAR}, where the connected open squares indicate the median
$g_{\rm bar}$-$g_{\rm tot}$ relation for all APOSTLE centrals. 
The thick dotted line follows Eq.~\ref{EqGtotGbar}, and it is clear
from the comparison that {\it isolated} APOSTLE galaxies follow a very
similar relation to the observed one, at least for
$g_{\rm bar}>10^{-12} {\rm \, m\, s}^{-2}$. At lower $g_{\rm bar}$ the
total accelerations of APOSTLE centrals approach $a_{\rm min}$.

Tidal stripping is expected to modify this relation, reducing
$g_{\rm bar}$ and shifting satellites to $g_{\rm tot}$ values well
below $a_{\rm min}$. This is illustrated by the coloured lines in the
right panel of Fig.~\ref{FigMDAR}, which indicate where faint
dwarfs affected by tidal stripping would be expected to lie, depending
on their half-mass radius. Satellites affected little by stripping
(shown in red) are expected to continue the flattening trend,
whereas heavily stripped satellites should
fall below the $a_{\rm min}$ boundary, and approach, in extreme cases
(shown in blue), the extrapolation of Eq.~\ref{EqGtotGbar} (dotted line).

A simple and robust prediction from APOSTLE-like models is then that
tidal stripping should scatter satellites below the mean
$g_{\rm bar}$-$g_{\rm tot}$ trend that holds for isolated systems,
leading to substantial spread in the value of $g_{\rm tot}$ at fixed
$g_{\rm bar}$ at the faint end.

This is, indeed, what is observed in the observational
data for LG satellite and field dwarfs. We show this in the
left-hand panel of Fig.~\ref{FigMDAR}, using for $g_{\rm tot}$ and
$g_{\rm bar}$ the values estimated at the half-mass radius, assuming
spherical symmetry for both the dark matter and baryonic
components, or,  more specifically,
\begin{equation}
g_{\rm tot}=V_{1/2}^2/r_{1/2}, \quad g_{\rm bar}=G\,\Mstr/2 \, r_{1/2}^2
\label{EqG}
\end{equation}

The data in this panel show that the tight MDAR reported by
\citet{McGaugh2016} and \citet{Lelli2016} for brighter galaxies breaks
down in the very faint, low-surface brightness regime 
($g_{\rm bar}<10^{-12} {\rm m\, s}^{-2}$). The scatter in $g_{\rm tot}$
at given $g_{bar}$
spreads nearly two decades, seriously calling into question the idea
that MDAR might encode a `natural law' that allows the total
gravitational acceleration to be accurately estimated from the
baryonic contribution alone.

The observed data, on the other hand, are quite consistent with the
APOSTLE predictions, once the effects of tidal stripping are taken
into account. Interestingly, our models predict that the most heavily
tidally stripped satellites should approach the extrapolation of
Eq.~\ref{EqGtotGbar}. (Cra~2 is one example of several in that
regard.) On the other hand, those largely unaffected by tides should
hover just above the $g_{\rm tot}=a_{\rm min}$ line, as observed. More
moderately stripped systems should bridge the gap between the two, just
as observed in the left-hand panel of Fig.~\ref{FigMDAR}.

We conclude that the overall behaviour of dwarf satellites galaxies in
the $g_{\rm obs}$ vs $g_{\rm bar}$ plane can be understood in the
$\Lambda$CDM framework as a simple consequence of tidal stripping.

\subsection{MOND and the velocity dispersion of LG dwarfs}
\label{SecMOND}

The extremely low accelerations of faint dwarfs lie in the regime
where the modified Newtonian gravity theory MOND \citep{Milgrom1983}
makes definite and clear predictions---the `deep-MOND limit'. In
this regime, the characteristic velocity of a non-rotating stellar
spheroid is determined solely by its mass (equal to that of the stellar
component in the case of a dSph) and by the MOND acceleration
parameter,
$a_0=1.2\times 10^{-10} {\rm m\, s}^{-2} =3.7 \times 10^3 {\rm
  \, km}^2 {\rm \, s}^{-2} {\rm \, kpc}^{-1}$ \citep{Milgrom2012a}.

Following \citet{McGaugh2013}, the MOND velocity dispersion may
be written as:
\begin{equation}
\sigma_{\rm iMOND} = (4G\Mstr \, a_{0}/81)^{1/4},
\label{EqSigmaiMond}
\end{equation}
where the `iMOND' subscript has been used to denote the fact that
this calculation assumes that the system is {\it isolated}
from more massive objects. 

MOND predictions for satellite galaxies are more uncertain, since they
are also subject to the external acceleration of their host,
$g_{\rm ex}=V_{\rm host}^2/D_{\rm host}$, where $V_{\rm host}$ is the
circular velocity of the host and $D_{\rm host}$ is the distance from
the satellite to the centre of the primary.  The MOND prediction is
modified by this `external field effect" (EFE), introducing a
correction to Eq.~\ref{EqSigmaiMond} whose importance will depend on
the ratio of `external' to `internal' acceleration for each dwarf.

Approximating the internal acceleration
by $g_{\rm in}=3\,\sigma_{\rm iMOND}^2/r_{1/2}$, it is possible to
compute the MOND prediction in the regime where $g_{\rm ex}\gg g_{\rm in}$.
In this case, the MOND velocity dispersion resembles our Eq.~\ref{EqV1/2}, but 
substituting the gravitational constant, $G$, by its `effective'
value at the location of the satellite,
$G_{\rm eff}\approx G\, a_0/g_{\rm ex}$. In other words, 
\begin{equation}
\sigma_{\rm eMOND}=(G_{\rm eff} \, \Mstr / r_{1/2})^{1/2}, \quad {\rm
  if}\, g_{\rm in} \ll g_{\rm ex}.
\label{EqSigmaEFE}
\end{equation}
Where `eMOND' refers to EFE dominance. We shall assume a constant
value of $V_{\rm host}=220 \kms$ and $230 \kms$ for the Milky Way and
M31 satellites, respectively.

We compare the isolated MOND predictions with LG dwarf data in the
left panel of Fig.~\ref{FigMOND}. Clearly, a number of dwarfs deviate
systematically from the MOND prediction, especially at the very faint
end, where the velocity dispersions of `ultra-faint' dwarfs exceed
the MOND predictions by a large factor.

Could this offset be caused by the `external field effect'? We
explore this in the right-hand panel of Fig.~\ref{FigMOND},
where we plot $\Mstr$ as a function of the ratio, $g_{\rm ex}/g_{\rm
  in}$ \footnote{For field dwarf galaxies, $g_{\rm ex}$ is estimated
  by considering the distance and $V_{\rm host}$ of the closest
  primary. Assuming a flat rotation curve for the host out to large
  distances overestimates $g_{\rm ex}$; hence the left-pointed arrow
  for field dwarfs on this plot.}. We can see that many of the
ultra-faint dwarfs where the iMOND prediction fails are indeed in a
regime where EFEs are dominant. Although the theory does not
specify precisely when the EFE formula (Eq.~\ref{EqSigmaEFE}) should
replace the isolated MOND prediction (Eq.~\ref{EqSigmaiMond}), we can
check at least whether EFE corrections are likely to help by comparing
the data with a weighted mean of the two:
\begin{equation}
\sigma_{\rm MOND}={g_{\rm in} \, \sigma_{\rm iMOND}+g_{\rm ex}\,
  \sigma_{\rm eMOND} \over g_{\rm in}+g_{\rm ex}}.
\label{EqSigmaMond}
\end{equation}

\begin{figure}
  \hspace{-0.2cm}
  \resizebox{8.5cm}{!}{\includegraphics{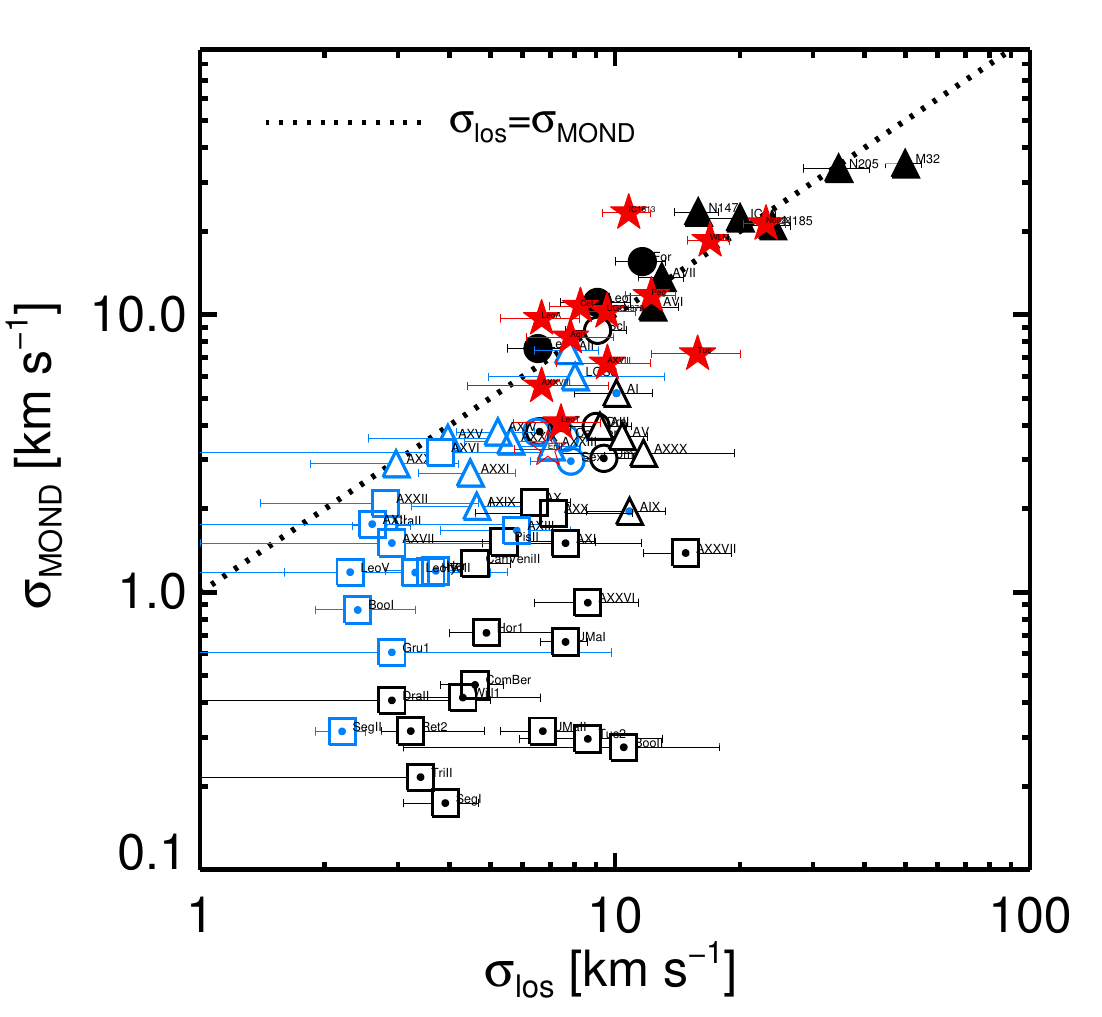}}\\%
  \caption{Velocity dispersion of LG dwarfs compared with MOND
    predictions and taking into account external field effects
    (Eq. \ref{EqSigmaMond}). Systems with $g_{\rm in}>g_{\rm ex}$ are
    shown with filled symbols; systems with $g_{\rm ex}>g_{\rm in}$
    are shown with open symbols. Those in the `EFE-dominated' regime
    ($g_{\rm ex}/g_{\rm in}>5$) are highlighted with a dot. Note that
    MOND clearly fails to account for the observed velocity
    dispersions of many LG dwarfs, especially those at the extremely
    faint end.}
\label{FigMOND2}
\end{figure}

We show the comparison in Fig.~\ref{FigMOND2}, where we compare
observed velocity dispersions with the predictions of
Eq.~\ref{EqSigmaMond}. Filled symbols in this figure identify systems
where $g_{\rm ex}<g_{\rm in}$; `dotted' symbols those in the
EFE-dominated regime $g_{\rm ex}>5\, g_{\rm in}$, and open symbols
those in the intermediate regime. As is clear from this figure, EFE
corrections actually make matters worse, as it predicts even lower
velocity dispersions than iMOND at the very faint end. We conclude
that MOND fails to account for the observed velocity
dispersions of LG dwarfs.
 
It is unclear how this result may be reconciled with MOND, but it adds
to a long list of observations where MOND encounters serious
difficulties, such as, for example, the centres of galaxy clusters
\citep{Gerbal1992,Sanders2003} and the properties of the Ly-$\alpha$
forest \citep{Aguirre2001}. What makes the result in
Fig.~\ref{FigMOND2} particularly compelling is that most of the dwarfs
in this graph are deep in the MOND regime, where the predictions of
the theory should be particularly reliable. We conclude that the
observed velocity dispersion of ultra-faint dwarfs pose a possibly
insurmountable challenge to that theory.

\section{Summary and Conclusions}
\label{SecConc}

The low velocity dispersions of dwarf galaxies have long been
difficult to reconcile with the $\Lambda$CDM standard model of
structure formation. This is because dwarfs in $\Lambda$CDM are
expected to form in haloes above a certain minimum circular velocity
of order $20$-$30 \kms$, which is at odds with the very low velocity
dispersions, $\sigma_{\rm los}\sim 3$--$5 \kms$,  of a number of Local
Group satellites.


Previously proposed solutions include the possibility that baryons may
have reduced the expected dark matter content of a dwarf by carving a
constant density `core' in the dark mass profile
\citep{DiCintio2014,Onorbe2015}, or, alternatively, that the stellar
component of dwarfs samples only the very inner, rising part of the
CDM circular velocity curve \citep{Benson2002,Stoehr2002}. The first
possibility has been hinted at by recent simulation work, but it is
unlikely to apply in the regime of extremely dark matter-dominated
ultra-faint dwarfs, where there are simply not enough baryons to
modify the dark matter profile.

The second possibility has been contradicted by the discovery of `cold
faint giants'; i.e., dwarfs that are exceptional because of their low
luminosity, large size and cold kinematics. Examples include Cra~2
and And~XIX, dwarfs so large that their stellar kinematics should
faithfully sample the maximum circular velocity of the halo, but whose
stars are kinematically much colder than expected.

We have examined here the possibility that these issues might
be solved by considering the effects of tidal stripping. Our analysis
uses the galaxy mass-halo mass relation from the APOSTLE cosmological
simulations of the Local Group, as well as guidance from earlier
N-body work about the changes induced by tidal stripping on the size,
stellar mass, and velocity dispersion of spheroidal galaxies embedded
in cuspy CDM haloes. Our main conclusions may be summarized as
follows.

\begin{itemize}

\item The APOSTLE simulations predict that all faint isolated dwarfs
  (i.e., $\Mstr<10^7\, M_\odot$) should inhabit haloes that span a
  fairly narrow range of virial masses. Together with the self-similar
  nature of CDM haloes, this implies a strong correlation between dwarf
  size and characteristic velocity, as larger galaxies should
  encompass more dark mass. Systems that fail to follow this expected
  correlation have likely been affected by tidal stripping.

\item Prior N-body work on the tidal evolution of dSphs in CDM haloes
  (PNM08 and E15) allows us to `undo' the effects of tides on the size, mass,
  and velocity dispersion of a satellite. The change in each of these
  parameters is linked to the others through `tidal tracks' that may
  be used to recover the original structural parameters of a
  satellite's progenitor. Importantly, these tracks suggest that the
  stellar half-mass radius of a satellite is the least affected by
  tides, even for cases of extreme mass loss.

\item Satellite progenitors, when constrained to match the APOSTLE
  $\Mstr$ vs $V_{\rm max}$ relation, follow scaling laws linking the
  stellar mass, size, and velocity dispersion that are in excellent
  agreement with those of isolated field galaxies. This provides an
  attractive explanation for (i) why the [Fe/H]-$\sigma$ relation
  flattens at low $\sigma$; (ii) why some faint satellites are
  extremely large (they are the tidal remnants of once more massive,
  intrinsically large galaxies), and (iii) why satellites and
  field dwarfs follow a similar dynamical mass-to-light ratio vs
  luminosity relation, regardless of stripping.

\item Tidally stripped satellites are closer than the average to the
  centre of their host, but even very highly tidally-stripped systems
  are found as out as $\sim 200$ kpc from the centre. We find no
  obvious inconsistency between the degree of tidal stripping
  predicted by our models and the measured spatial distribution of LG
  satellites.

\item Tidal stripping is expected to result in large scatter at the
  faint, low-acceleration end of the mass discrepancy-acceleration
  relation (MDAR) that holds for brighter late-type
  galaxies. Satellites that have lost substantial amounts of dark
  matter to tides are pushed to accelerations well below the nominal
  minimum, $a_{\rm min}=10^{-11} {\rm m\, s}^{-2}$, expected for
  isolated dwarfs in $\Lambda$CDM. The expected scatter is
  consistent with LG dwarf observations, but inconsistent with the
  idea that a single MDAR relation holds for all galaxies.

\item Finally, the low velocity dispersion population of satellites is
  plainly inconsistent with the predictions of Modified Newtonian
  Dynamics: MOND predicts, at the very faint end, much lower velocity
  dispersions than observed. Resorting to `external field effects'
  induced by the primaries does not help, and actually makes MOND
  predictions even more inconsistent with extant data. The velocity
  dispersions of the faintest galaxies known might prove an
  insurmountable difficulty for this theory.

\end{itemize}

Although appealing as a scenario, our proposal that tidal stripping
might help to reconcile the peculiar properties of a number of
satellites with the predictions of $\Lambda$CDM has a number of
potential problems that need to be fully addressed in future work and
that would also benefit from insight from other cosmological
simulations of Local Group environments \citep[see;
e.g.,][]{Wang2015a,Wetzel2016}.  One potentially weak point concerns the relatively
high frequency of highly stripped LG satellites required to match the
LG dwarfs. Indeed, we find that about $\sim 11$ ($16$) per cent of MW
and M31 satellites brighter than $M_V=-8$ ($-5.5$) have lost more than
$90$ per cent of their original stellar mass. Unfortunately, current
APOSTLE simulations do not have adequate numerical resolution to make
accurate predictions that may be compared with these data. This is an
issue, however, that should be revisited with simulations of higher
resolution, as they become available.

A further, related point is that a number of dwarfs are deemed to have
undergone rather dramatic transformation because of tides. Cra~2, And
XIX, And~XXI, And~XXV, and Bootes~1, for example, would all need to
have shed roughly $99$ per cent of their original mass for their
progenitors to be consistent with APOSTLE dwarfs, yet there is little
evidence in the galaxies themselves or in their surroundings for this
rather extreme mass loss. Simulations, however, make
  some fairly robust predictions for these heavily-stripped satellites
  that may be constrasted with observation. Because they have been so
  heavily shaken by tides, we expect them their remnants to be round
  and their surface brightness profiles to have large King
  concentration values. In addition, because they have been stripped
  of their surrounding halos, their maximum circular velocities must
  be very similar to that inferred within their stellar half-mass
  radius, a prediction that may in principle be tested with accurate
  dynamical modeling of kinematic data.

Of course, identifying and quantifying debris from such events in the
halo of the Milky Way that may be traced back to these satellites
would also be an important step towards turning our proposal from
informed conjecture into a compelling picture. We
  anticipate, however, that this task will be rather difficult, given
  the extremely low surface brightness expected for the stream
  (fainter than the bound remnants, some of which are already at the
  limit of detectability). Another possibility would be to look for
  loosely bound stars in the immediate vicinity of the tidally
  affected dwarf, which would flatten the satellite surface density
  profile outside a characteristic radius
  \citep{Penarrubia2009}. Detecting such stars would also be extremely
  challenging, since simulations indicate that their surface
  brightness, at apocentre, might be up to $\sim 10$ magnitudes
  fainter than the central surface density of the satellite
  \citep[see, e.g.,][]{Tomozeiu2016}.

Proper motions of individual stars would be of immense help. These
could be used to estimate pericentric distances and orbital times that
may be used to check the consistency of our model with more detailed
modelling of each individual system suspected to be a `tidal remnant'.
We very much look forward to such data in order to inform our analysis
further in future work.

\section{Acknowledgements}

We acknowledge useful discussions with Alan McConnachie and Joop Schaye. We
are thankful to Marla Geha, Erik Tollerud, and Ryan Leaman for sharing
data with us. The research was supported in part by the Science and
Technology Facilities Council Consolidated Grant (ST/P000541/1), and
the European Research Council under the European Union Seventh
Framework Programme (FP7/2007-2013)/ERC Grant agreement
278594-GasAroundGalaxies. CSF acknowledges ERC Advanced Grant 267291
COSMIWAY; and JW the 973 program grant 2015CB857005 and NSFC grant
No. 11373029, 11390372. This work used the DiRAC Data Centric system
at Durham University, operated by the Institute for Computational
Cosmology on behalf of the STFC DiRAC HPC Facility (www.dirac.ac.uk),
and also resources provided by WestGrid (www.westgrid.ca) and Compute
Canada (www.computecanada.ca). The DiRAC system was funded by BIS
National E-infrastructure capital grant ST/K00042X/1, STFC capital
grants ST/H008519/1 and ST/K00087X/1, STFC DiRAC Operations grant
ST/K003267/1 and Durham University. DiRAC is part of the National
E-Infrastructure. This research has made use of NASA's Astrophysics
Data System.


\bibliographystyle{mnras}
\bibliography{master}
\bsp

\appendix

\section{Parameters for dwarf galaxies in the Local Group}
Table of the observed parameters of dwarf galaxies in the Local Group,
adopted in this work, along with tables containing derived parameters.

\onecolumn
\def\arraystretch{1.3}
\begin{ThreePartTable}
\begin{TableNotes}
\footnotesize
\item {\bf References}: Most parameters are adopted from the updated
  (October 2015) version of the tables from
  \citet{McConnachie2012}. We also use parameters from other
  references, including the following: 1: \citet{McConnachie2012}, 2: \citet{Torrealba2016},
  3: \citet{Caldwell2017}, 4: \citet{Koposov2015b},
  5:\citet{Walker2016}, 6: \citet{Martin2016b}, 7: \citet{Kirby2015},
  8: \citet{Kirby2017}, 9: \citet{Tollerud2012}, 10: \citet{Ho2012},
  11: \citet{Collins2013}, 12: \citet{Tollerud2013}, 13:
  \citet{Collins2010}, 14: \citet{Martin2016}, 15:
  \citet{Kirby2014}, 16: \citet{Hunter2006}, 17: \citet{Leaman2012},
  18: \citet{Bellazzini2011}, 19: \citet{Kirby2017a}, 20:
  \citet{McConnachie2006b}, 21: \citet{Crnojevic2016}, 22: \citet{Li2017}.
\end{TableNotes}

\begin{longtable}{{l} *{7}{c} {r}}
  \caption{Observed parameters of dwarf galaxies in the Local Group}\\
  \hline
  \hline
  Gal. Name  &  $m_V$ & $(m-M)_0$ &  $R_{\rm eff}$ & $\sigma_{\rm los}$ &  $\Mstr/L_V$ &  [Fe/H] & $D_{\rm host}$ & references \\
  & & & (arcmin) & ($\kms$) & & (dex) & (kpc) \\
  \endfirsthead

  {{\bfseries \tablename\ \thetable{} -- (continued)}} \\
  \hline
  \hline
  Gal. Name  &  $m_V$ & $(m-M)_0$ &  $R_{\rm eff}$ & $\sigma_{\rm los}$ &  $\Mstr/L_V$ &  [Fe/H] & $D_{\rm host}$ & references\\
  & & & arcmin & $\kms$ & & dex & kpc \\
  \hline
  \endhead

  \insertTableNotes  
  \endlastfoot
  
  \hline

  \underline{{\bf MW satellites}} \\
       For & $   7.4^{+ 0.3 }_{- 0.3 }$ & $ 20.84  ^{+0.18 }_{-0.18 }$ & $  16.6  ^{+ 1.2 }_{- 1.2 }$ & $  11.7  ^{+ 0.9 }_{- 0.9 }$ & $ 1.2  $ & $-0.90 \pm  0.01  $ & $149  $ &      1 \\
      LeoI & $  10.0^{+ 0.3 }_{- 0.3 }$ & $ 22.02  ^{+0.13 }_{-0.13 }$ & $   3.4  ^{+ 0.3 }_{- 0.3 }$ & $   9.2  ^{+ 1.4 }_{- 1.4 }$ & $ 0.9  $ & $-1.40 \pm  0.01  $ & $256  $ &      1 \\
       Scl & $   8.6^{+ 0.5 }_{- 0.5 }$ & $ 19.67  ^{+0.14 }_{-0.14 }$ & $  11.3  ^{+ 1.6 }_{- 1.6 }$ & $   9.2  ^{+ 1.1 }_{- 1.1 }$ & $ 1.7  $ & $-1.60 \pm  0.01  $ & $ 85  $ &      1 \\
     LeoII & $  12.0^{+ 0.3 }_{- 0.3 }$ & $ 21.84  ^{+0.13 }_{-0.13 }$ & $   2.6  ^{+ 0.6 }_{- 0.6 }$ & $   6.6  ^{+ 0.7 }_{- 0.7 }$ & $ 1.6  $ & $-1.60 \pm  0.01  $ & $235  $ &      1 \\
      SexI & $  10.4^{+ 0.5 }_{- 0.5 }$ & $ 19.67  ^{+0.10 }_{-0.10 }$ & $  27.8  ^{+ 1.2 }_{- 1.2 }$ & $   7.9  ^{+ 1.3 }_{- 1.3 }$ & $ 1.6  $ & $-1.90 \pm  0.01  $ & $ 88  $ &      1 \\
       Car & $  11.0^{+ 0.5 }_{- 0.5 }$ & $ 20.11  ^{+0.13 }_{-0.13 }$ & $   8.2  ^{+ 1.2 }_{- 1.2 }$ & $   6.6  ^{+ 1.2 }_{- 1.2 }$ & $ 1.0  $ & $-1.70 \pm  0.01  $ & $106  $ &      1 \\
       Dra & $  10.6^{+ 0.2 }_{- 0.2 }$ & $ 19.40  ^{+0.17 }_{-0.17 }$ & $  10.0  ^{+ 0.3 }_{- 0.3 }$ & $   9.1  ^{+ 1.2 }_{- 1.2 }$ & $ 1.8  $ & $-1.90 \pm  0.01  $ & $ 75  $ &      1 \\
       Umi & $  10.6^{+ 0.5 }_{- 0.5 }$ & $ 19.40  ^{+0.10 }_{-0.10 }$ & $  19.9  ^{+ 1.9 }_{- 1.9 }$ & $   9.5  ^{+ 1.2 }_{- 1.2 }$ & $ 1.9  $ & $-2.10 \pm  0.01  $ & $ 77  $ &     23 \\
   CanVenI & $  13.1^{+ 0.2 }_{- 0.2 }$ & $ 21.69  ^{+0.10 }_{-0.10 }$ & $   8.9  ^{+ 0.4 }_{- 0.4 }$ & $   7.6  ^{+ 0.4 }_{- 0.4 }$ & $ 1.6  $ & $-1.90 \pm  0.01  $ & $216  $ &      1 \\
     CraII & $  12.1^{+ 0.1 }_{- 0.1 }$ & $ 20.35  ^{+0.02 }_{-0.02 }$ & $  29.2  ^{+ 2.4 }_{- 2.2 }$ & $   2.8  ^{+ 0.3 }_{- 0.3 }$ & $ 1.6  $ & $-1.98 \pm  0.10  $ & $114  $ &    2,3 \\
       Her & $  14.0  ^{+ 0.3 }_{- 0.3 }$ & $ 20.60  ^{+0.20 }_{-0.20 }$ & $   8.6  ^{+ 1.8 }_{- 1.1 }$ & $   3.7  ^{+ 0.9 }_{- 0.9 }$ & $ 1.6 $ & $-2.40  \pm  0.04  $ & $125 $ &          1 \\
      BooI & $  12.8  ^{+ 0.2 }_{- 0.2 }$ & $ 19.11  ^{+0.08 }_{-0.08 }$ & $  12.6  ^{+ 1.0 }_{- 1.0 }$ & $   2.4  ^{+ 0.5 }_{- 0.9 }$ & $ 1.6 $ & $-2.50  \pm  0.11  $ & $ 63 $ &          1 \\
     LeoIV & $  15.1  ^{+ 0.4 }_{- 0.4 }$ & $ 20.94  ^{+0.09 }_{-0.09 }$ & $   4.6  ^{+ 0.8 }_{- 0.8 }$ & $   3.3  ^{+ 1.7 }_{- 1.7 }$ & $ 1.6 $ & $-2.50  \pm  0.07  $ & $154 $ &          1 \\
      UMaI & $  14.4  ^{+ 0.3 }_{- 0.3 }$ & $ 19.93  ^{+0.10 }_{-0.10 }$ & $  11.3  ^{+ 1.7 }_{- 1.7 }$ & $   7.6  ^{+ 1.0 }_{- 1.0 }$ & $ 1.6 $ & $-2.10  \pm  0.04  $ & $100 $ &          1 \\
      LeoV & $  16.0  ^{+ 0.4 }_{- 0.4 }$ & $ 21.25  ^{+0.12 }_{-0.12 }$ & $   2.6  ^{+ 0.6 }_{- 0.6 }$ & $   2.3  ^{+ 1.6 }_{- 3.2 }$ & $ 1.6 $ & $-2.00  \pm  0.20  $ & $177 $ &          1 \\
     PisII & $  16.3  ^{+ 0.5 }_{- 0.5 }$ & $ 21.30  ^{+0.50 }_{-0.50 }$ & $   1.1  ^{+ 0.1 }_{- 0.1 }$ & $   5.4  ^{+ 2.4 }_{- 3.6 }$ & $ 1.6 $ & $-2.45  \pm  0.07  $ & $180 $ &          7 \\
 CanVeniII & $  16.1  ^{+ 0.5 }_{- 0.5 }$ & $ 21.02  ^{+0.06 }_{-0.06 }$ & $   1.6  ^{+ 0.3 }_{- 0.3 }$ & $   4.6  ^{+ 1.0 }_{- 1.0 }$ & $ 1.6 $ & $-2.20  \pm  0.05  $ & $159 $ &          1 \\
     HydII & $  15.8  ^{+ 0.3 }_{- 0.3 }$ & $ 20.64  ^{+0.16 }_{-0.16 }$ & $   1.7  ^{+ 0.3 }_{- 0.2 }$ & $   <3.6$                  & $ 1.6 $ & $-2.00  \pm  0.08  $ & $131 $ &          7 \\
     UMaII & $  13.3  ^{+ 0.5 }_{- 0.5 }$ & $ 17.50  ^{+0.30 }_{-0.30 }$ & $  16.0  ^{+ 1.0 }_{- 1.0 }$ & $   6.7  ^{+ 1.4 }_{- 1.4 }$ & $ 1.6 $ & $-2.40  \pm  0.06  $ & $ 37 $ &          1 \\
    ComBer & $  14.1  ^{+ 0.5 }_{- 0.5 }$ & $ 18.20  ^{+0.20 }_{-0.20 }$ & $   6.0  ^{+ 0.6 }_{- 0.6 }$ & $   4.6  ^{+ 0.8 }_{- 0.8 }$ & $ 1.6 $ & $-2.60  \pm  0.05  $ & $ 44 $ &          1 \\
      Tuc2 & $  15.0  ^{+ 0.1 }_{- 0.1 }$ & $ 18.80  ^{+0.20 }_{-0.20 }$ & $   9.8  ^{+ 1.7 }_{- 1.1 }$ & $   8.6  ^{+ 2.7 }_{- 4.4 }$ & $ 1.6 $ & $-2.23  \pm  0.15  $ & $ 53 $ &          5 \\
      Hor1 & $  16.1  ^{+ 0.1 }_{- 0.1 }$ & $ 19.50  ^{+0.20 }_{-0.20 }$ & $   1.3  ^{+ 0.2 }_{- 0.1 }$ & $   4.9  ^{+ 0.9 }_{- 2.8 }$ & $ 1.6 $ & $-2.76  \pm  0.17  $ & $ 79 $ &          4 \\
      Gru1 & $  17.0  ^{+ 0.3 }_{- 0.3 }$ & $ 20.40  ^{+0.20 }_{-0.20 }$ & $   1.8  ^{+ 0.9 }_{- 0.4 }$ & $   2.9  ^{+ 2.1 }_{- 6.9 }$ & $ 1.6 $ & $-1.42  \pm  0.50  $ & $116 $ &          5 \\
     DraII & $  14.0  ^{+ 0.8 }_{- 0.8 }$ & $ 16.90  ^{+0.30 }_{-0.30 }$ & $   2.7  ^{+ 1.0 }_{- 0.8 }$ & $   2.9  ^{+ 2.1 }_{- 2.1 }$ & $ 1.6 $ & $-2.20    $ & $ 25 $ &          6 \\
     BooII & $  15.4  ^{+ 0.9 }_{- 0.9 }$ & $ 18.10  ^{+0.06 }_{-0.06 }$ & $   4.2  ^{+ 1.4 }_{- 1.4 }$ & $  10.5  ^{+ 7.4 }_{- 7.4 }$ & $ 1.6 $ & $-1.70  \pm  0.05  $ & $ 38 $ &          1 \\
      Ret2 & $  14.7  ^{+ 0.1 }_{- 0.1 }$ & $ 17.40  ^{+0.20 }_{-0.20 }$ & $   3.6  ^{+ 0.2 }_{- 0.1 }$ & $   3.2  ^{+ 0.5 }_{- 1.6 }$ & $ 1.6 $ & $-2.46  \pm  0.30  $ & $ 31 $ &          4 \\
     Will1 & $  15.2  ^{+ 0.7 }_{- 0.7 }$ & $ 17.90  ^{+0.40 }_{-0.40 }$ & $   2.3  ^{+ 0.4 }_{- 0.4 }$ & $   4.3  ^{+ 1.3 }_{- 2.3 }$ & $ 1.6 $ & $-2.10    $ & $ 42 $ &          1 \\
     SegII & $  15.2  ^{+ 0.3 }_{- 0.3 }$ & $ 17.70  ^{+0.10 }_{-0.10 }$ & $   3.4  ^{+ 0.2 }_{- 0.2 }$ & $   2.2  ^{+ 0.3 }_{- 0.3 }$ & $ 1.6 $ & $-2.20  \pm  0.13  $ & $ 40 $ &          1 \\
     TriII & $  15.6  ^{+ 0.5 }_{- 0.5 }$ & $ 17.40  ^{+0.10 }_{-0.10 }$ & $   3.9  ^{+ 1.1 }_{- 0.9 }$ & $   <3.4$                  & $ 1.6 $ & $-2.50   $ & $ 36 $ &          8 \\
      SegI & $  15.3  ^{+ 0.8 }_{- 0.8 }$ & $ 16.80  ^{+0.20 }_{-0.20 }$ & $   4.4  ^{+ 1.2 }_{- 0.6 }$ & $   3.9  ^{+ 0.8 }_{- 0.8 }$ & $ 1.6 $ & $-2.70  \pm  0.40  $ & $ 27 $ &          1 \\
  & \\
\underline{{\bf M31 satellites}} \\
      N205 & $   8.1  ^{+ 0.1 }_{- 0.1 }$ & $ 24.58  ^{+0.07 }_{-0.07 }$ & $   2.5  ^{+ 0.1 }_{- 0.1 }$ & $  35.0  ^{+ 5.0 }_{- 5.0 }$ & $ 1.4  $ & $-0.80 \pm  0.20  $ & $ 41  $ &         1 \\
       M32 & $   8.1  ^{+ 0.1 }_{- 0.1 }$ & $ 24.53  ^{+0.21 }_{-0.21 }$ & $   0.5  ^{+ 0.1 }_{- 0.1 }$ & $  50.0  ^{+ 0.0 }_{- 0.0 }$ & $ 1.6  $ & $-0.20   $ & $ 22  $ &         1 \\
      N185 & $   9.2  ^{+ 0.1 }_{- 0.1 }$ & $ 23.95  ^{+0.09 }_{-0.09 }$ & $   1.5  ^{+ 0.0 }_{- 0.0 }$ & $  24.0  ^{+ 1.0 }_{- 1.0 }$ & $ 1.0  $ & $-1.30 \pm  0.10  $ & $187  $ &         1 \\
      N147 & $   9.5  ^{+ 0.1 }_{- 0.1 }$ & $ 24.15  ^{+0.09 }_{-0.09 }$ & $   2.0  ^{+ 0.0 }_{- 0.0 }$ & $  16.0  ^{+ 1.0 }_{- 1.0 }$ & $ 1.6  $ & $-1.10 \pm  0.10  $ & $142  $ &         1 \\
      AVII & $  11.2  ^{+ 0.3 }_{- 0.3 }$ & $ 24.41  ^{+0.10 }_{-0.10 }$ & $   3.5  ^{+ 0.1 }_{- 0.1 }$ & $  13.0  ^{+ 1.0 }_{- 1.0 }$ & $ 0.9  $ & $-1.40 \pm  0.30  $ & $218  $ &         9 \\
       AII & $  11.5  ^{+ 0.2 }_{- 0.2 }$ & $ 24.07  ^{+0.06 }_{-0.06 }$ & $   6.2  ^{+ 0.2 }_{- 0.2 }$ & $   7.8  ^{+ 1.1 }_{- 1.1 }$ & $ 1.0  $ & $-1.30 \pm  0.03  $ & $184  $ &     10,14 \\
        AI & $  12.5  ^{+ 0.1 }_{- 0.1 }$ & $ 24.36  ^{+0.07 }_{-0.07 }$ & $   3.1  ^{+ 0.3 }_{- 0.3 }$ & $  10.2  ^{+ 1.9 }_{- 1.9 }$ & $ 1.6  $ & $-1.40 \pm  0.04  $ & $ 58  $ &      9,14 \\
       AVI & $  13.0  ^{+ 0.2 }_{- 0.2 }$ & $ 24.47  ^{+0.07 }_{-0.07 }$ & $   2.3  ^{+ 0.2 }_{- 0.2 }$ & $  12.4  ^{+ 1.5 }_{- 1.3 }$ & $ 1.6  $ & $-1.50 \pm  0.10  $ & $268  $ &        11 \\
    AXXIII & $  14.2  ^{+ 0.5 }_{- 0.5 }$ & $ 24.43  ^{+0.13 }_{-0.13 }$ & $   4.6  ^{+ 0.2 }_{- 0.2 }$ & $   7.1  ^{+ 1.0 }_{- 1.0 }$ & $ 1.6  $ & $-2.20 \pm  0.30  $ & $126  $ &     11,14 \\
      AIII & $  14.2  ^{+ 0.3 }_{- 0.3 }$ & $ 24.37  ^{+0.07 }_{-0.07 }$ & $   2.2  ^{+ 0.2 }_{- 0.2 }$ & $   9.3  ^{+ 1.4 }_{- 1.4 }$ & $ 1.8  $ & $-1.70 \pm  0.04  $ & $ 75  $ &      9,14 \\
      LGS3 & $  14.3  ^{+ 0.1 }_{- 0.1 }$ & $ 24.43  ^{+0.07 }_{-0.07 }$ & $   2.1  ^{+ 0.2 }_{- 0.2 }$ & $   7.9  ^{+ 5.3 }_{- 2.9 }$ & $ 1.0  $ & $-2.10 \pm  0.22  $ & $268  $ &         1 \\
      AXXI & $  14.8  ^{+ 0.6 }_{- 0.6 }$ & $ 24.59  ^{+0.06 }_{-0.07 }$ & $   3.5  ^{+ 0.3 }_{- 0.3 }$ & $   4.5  ^{+ 1.2 }_{- 1.0 }$ & $ 1.6  $ & $-1.80 \pm  0.10  $ & $133  $ &     11,14 \\
      AXXV & $  14.8  ^{+ 0.5 }_{- 0.5 }$ & $ 24.55  ^{+0.12 }_{-0.12 }$ & $   3.0  ^{+ 0.2 }_{- 0.2 }$ & $   3.0  ^{+ 1.2 }_{- 1.1 }$ & $ 1.6  $ & $-1.90 \pm  0.10  $ & $ 88  $ &     11,14 \\
        AV & $  14.9  ^{+ 0.2 }_{- 0.2 }$ & $ 24.44  ^{+0.08 }_{-0.08 }$ & $   1.4  ^{+ 0.2 }_{- 0.2 }$ & $  10.5  ^{+ 1.1 }_{- 1.1 }$ & $ 1.1  $ & $-2.00 \pm  0.10  $ & $109  $ &      9,14 \\
       AXV & $  14.6  ^{+ 0.3 }_{- 0.3 }$ & $ 23.98  ^{+0.26 }_{-0.12 }$ & $   1.2  ^{+ 0.1 }_{- 0.1 }$ & $   4.0  ^{+ 1.4 }_{- 1.4 }$ & $ 1.6  $ & $-1.80 \pm  0.20  $ & $178  $ &      9,14 \\
      AXIX & $  15.6  ^{+ 0.6 }_{- 0.6 }$ & $ 24.57  ^{+0.08 }_{-0.43 }$ & $   6.2  ^{+ 0.1 }_{- 0.1 }$ & $   4.7  ^{+ 1.6 }_{- 1.4 }$ & $ 1.6  $ & $-1.80 \pm  0.30  $ & $113  $ &     11,14 \\
      AXIV & $  15.9  ^{+ 0.5 }_{- 0.5 }$ & $ 24.50  ^{+0.06 }_{-0.56 }$ & $   1.7  ^{+ 0.8 }_{- 0.8 }$ & $   5.3  ^{+ 1.0 }_{- 1.0 }$ & $ 1.6  $ & $-2.20 \pm  0.05  $ & $161  $ &      9,14 \\
     AXXIX & $  16.0  ^{+ 0.4 }_{- 0.4 }$ & $ 24.32  ^{+0.22 }_{-0.22 }$ & $   1.7  ^{+ 0.2 }_{- 0.2 }$ & $   5.7  ^{+ 1.2 }_{- 1.2 }$ & $ 1.6  $ & $-1.80   $ & $188  $ &        12 \\
       AIX & $  16.3  ^{+ 1.1 }_{- 1.1 }$ & $ 24.42  ^{+0.07 }_{-0.07 }$ & $   2.5  ^{+ 0.1 }_{- 0.1 }$ & $  10.9  ^{+ 2.0 }_{- 2.0 }$ & $ 1.6  $ & $-2.20 \pm  0.20  $ & $ 40  $ &      9,14 \\
      AXXX & $  16.2  ^{+ 0.3 }_{- 0.3 }$ & $ 24.17  ^{+0.10 }_{-0.26 }$ & $   1.4  ^{+ 0.1 }_{- 0.2 }$ & $  11.8  ^{+ 7.7 }_{- 4.7 }$ & $ 1.6  $ & $-1.70 \pm  0.40  $ & $147  $ &     11,14 \\
    AXXVII & $  16.7  ^{+ 0.5 }_{- 0.7 }$ & $ 24.59  ^{+0.12 }_{-0.12 }$ & $   1.8  ^{+ 0.3 }_{- 0.3 }$ & $  14.8  ^{+ 3.1 }_{- 4.3 }$ & $ 1.6 $ & $-2.10  \pm  0.50  $ & $ 73 $ &      11,14 \\
     AXVII & $  16.6  ^{+ 0.3 }_{- 0.3 }$ & $ 24.31  ^{+0.11 }_{-0.08 }$ & $   1.4  ^{+ 0.3 }_{- 0.3 }$ & $   2.9  ^{+ 1.9 }_{- 2.2 }$ & $ 1.6 $ & $-1.70  \pm  0.20  $ & $ 70 $ &      11,14 \\
        AX & $  16.7  ^{+ 0.3 }_{- 0.3 }$ & $ 24.13  ^{+0.08 }_{-0.13 }$ & $   1.1  ^{+ 0.4 }_{- 0.2 }$ & $   6.4  ^{+ 1.4 }_{- 1.4 }$ & $ 1.6 $ & $-1.90  \pm  0.11  $ & $134 $ &       9,14 \\
      AXVI & $  16.1  ^{+ 0.3 }_{- 0.3 }$ & $ 23.39  ^{+0.19 }_{-0.14 }$ & $   1.0  ^{+ 0.1 }_{- 0.1 }$ & $   3.8  ^{+ 2.9 }_{- 2.9 }$ & $ 1.6 $ & $-2.10  \pm  0.20  $ & $323 $ &       9,14 \\
      AXII & $  17.7  ^{+ 0.5 }_{- 0.5 }$ & $ 24.70  ^{+0.30 }_{-0.30 }$ & $   1.1  ^{+ 0.2 }_{- 0.2 }$ & $   2.6  ^{+ 2.6 }_{- 5.1 }$ & $ 1.6 $ & $-2.20  \pm  0.20  $ & $178 $ &      13,14 \\
     AXIII & $  17.8  ^{+ 0.4 }_{- 0.4 }$ & $ 24.62  ^{+0.05 }_{-0.05 }$ & $   0.8  ^{+ 0.4 }_{- 0.3 }$ & $   5.8  ^{+ 2.0 }_{- 2.0 }$ & $ 1.6 $ & $-1.70  \pm  0.30  $ & $132 $ &       9,14 \\
     AXXII & $  18.0  ^{+ 0.4 }_{- 0.4 }$ & $ 24.82  ^{+0.07 }_{-0.07 }$ & $   0.9  ^{+ 0.3 }_{- 0.2 }$ & $   2.8  ^{+ 1.4 }_{- 1.9 }$ & $ 1.6 $ & $-1.80  \pm  0.60  $ & $273 $ &      11,14 \\
       AXX & $  18.0  ^{+ 0.4 }_{- 0.4 }$ & $ 24.35  ^{+0.12 }_{-0.12 }$ & $   0.4  ^{+ 0.2 }_{- 0.1 }$ & $   7.1  ^{+ 2.5 }_{- 3.9 }$ & $ 1.6 $ & $-2.20  \pm  0.40  $ & $129 $ &      11,14 \\
       AXI & $  18.0  ^{+ 0.4 }_{- 0.4 }$ & $ 24.33  ^{+0.05 }_{-0.05 }$ & $   0.6  ^{+ 0.2 }_{- 0.2 }$ & $   7.6  ^{+ 2.8 }_{- 4.0 }$ & $ 1.6 $ & $-1.80  \pm  0.10  $ & $110 $ &      11,14 \\
     AXXVI & $  18.5  ^{+ 0.7 }_{- 0.5 }$ & $ 24.41  ^{+0.12 }_{-0.12 }$ & $   1.0  ^{+ 0.6 }_{- 0.5 }$ & $   8.6  ^{+ 2.2 }_{- 2.8 }$ & $ 1.6 $ & $-1.80  \pm  0.50  $ & $102 $ &      11,14 \\
 & \\
\underline{{\bf LG field dwarfs}} \\
     N6822 & $   8.1  ^{+ 0.2 }_{- 0.2 }$ & $ 23.31  ^{+0.08 }_{-0.08 }$ & $   3.6  ^{+ 0.2 }_{- 0.2 }$ & $  23.2  ^{+ 1.2 }_{- 1.2 }$ & $ 0.8  $ & $-1.00 \pm  0.50  $ & $ 451 $ &      15,16 \\
    IC1613 & $   9.2  ^{+ 0.1 }_{- 0.1 }$ & $ 24.39  ^{+0.12 }_{-0.12 }$ & $   4.7  ^{+ 0.4 }_{- 0.4 }$ & $  10.8  ^{+ 1.0 }_{- 0.9 }$ & $ 1.0  $ & $-1.60 \pm  0.20  $ & $ 757 $ &      15,16 \\
       WLM & $  10.6  ^{+ 0.1 }_{- 0.1 }$ & $ 24.85  ^{+0.08 }_{-0.08 }$ & $   5.8  ^{+ 0.4 }_{- 0.3 }$ & $  17.0  ^{+ 1.0 }_{- 1.0 }$ & $ 0.9  $ & $-1.20 \pm  0.02  $ & $ 932 $ &         17 \\
   UGC4879 & $  13.2  ^{+ 0.2 }_{- 0.2 }$ & $ 25.67  ^{+0.04 }_{-0.04 }$ & $   0.40  ^{+ 0.04 }_{- 0.04 }$ & $   9.6  ^{+ 1.3 }_{- 1.2 }$ & $ 0.7  $ & $-1.50 \pm  0.06  $ & $1367 $ &      15,18 \\
       Peg & $  12.6  ^{+ 0.2 }_{- 0.2 }$ & $ 24.82  ^{+0.07 }_{-0.07 }$ & $   2.6  ^{+ 0.2 }_{- 0.2 }$ & $  12.3  ^{+ 1.2 }_{- 1.1 }$ & $ 1.0  $ & $-1.40 \pm  0.20  $ & $ 920 $ &      15,16 \\
      LeoA & $  12.4  ^{+ 0.2 }_{- 0.2 }$ & $ 24.51  ^{+0.12 }_{-0.12 }$ & $   1.5  ^{+ 0.1 }_{- 0.1 }$ & $   6.7  ^{+ 1.4 }_{- 1.2 }$ & $ 0.5  $ & $-1.40 \pm  0.20  $ & $ 801 $ &      19,16 \\
       Cet & $  13.1  ^{+ 0.2 }_{- 0.2 }$ & $ 24.39  ^{+0.07 }_{-0.07 }$ & $   3.2  ^{+ 0.1 }_{- 0.1 }$ & $   8.3  ^{+ 1.0 }_{- 1.0 }$ & $ 1.6  $ & $-1.90 \pm  0.10  $ & $ 755 $ &      20,16 \\
       Aqu & $  14.5  ^{+ 0.1 }_{- 0.1 }$ & $ 25.15  ^{+0.08 }_{-0.08 }$ & $   1.5  ^{+ 0.04 }_{- 0.04 }$ & $   7.9  ^{+ 1.9 }_{- 1.6 }$ & $ 1.0  $ & $-1.30 \pm  0.20  $ & $1065 $ &      20,16 \\
       Tuc & $  15.2  ^{+ 0.2 }_{- 0.2 }$ & $ 24.74  ^{+0.12 }_{-0.12 }$ & $   1.1  ^{+ 0.0 }_{- 0.0 }$ & $  15.8  ^{+ 4.1 }_{- 3.1 }$ & $ 1.6  $ & $-1.90 \pm  0.15  $ & $ 882 $ &          1 \\
    AXVIII & $  16.0  ^{+ 0.2 }_{- 0.2 }$ & $ 25.42  ^{+0.07 }_{-0.08 }$ & $   0.9  ^{+ 0.1 }_{- 0.1 }$ & $   9.7  ^{+ 2.3 }_{- 2.3 }$ & $ 1.6  $ & $-1.40 \pm  0.30  $ & $1216 $ &          9 \\
   AXXVIII & $  15.6  ^{+ 0.4 }_{- 0.9 }$ & $ 24.10  ^{+0.50 }_{-0.20 }$ & $   1.1  ^{+ 0.2 }_{- 0.2 }$ & $   6.6  ^{+ 2.9 }_{- 2.1 }$ & $ 1.6  $ & $-2.10 \pm  0.30  $ & $ 660 $ &         11 \\
      LeoT & $  15.1  ^{+ 0.5 }_{- 0.5 }$ & $ 23.10  ^{+0.10 }_{-0.10 }$ & $   1.0  ^{+ 0.1 }_{- 0.1 }$ & $   7.5  ^{+ 1.6 }_{- 1.6 }$ & $ 0.8  $ & $-2.00 \pm  0.05  $ & $ 421 $ &          1 \\
     EriII & $  15.7  ^{+ 0.2 }_{- 0.2 }$ & $ 22.80  ^{+0.10 }_{-0.10 }$ & $   2.3  ^{+ 0.1 }_{- 0.1 }$ & $   6.9  ^{+ 1.2 }_{- 0.9 }$ & $ 1.6  $ & $-2.38 \pm  0.13  $ & $ 381 $ &      21,22 \\

  \hline
  \label{TabData1}
\end{longtable}
\end{ThreePartTable}

\newpage

\onecolumn
\def\arraystretch{1.3}
\begin{longtable}{p{1cm} ccc p{1.2cm} p{1.2cm} cc p{1.2cm} p{1.2cm}}
  \caption{Derived parameters for dwarf galaxies in the Local Group}\\

  \hline
  \hline
  Gal. Name & $\Mstr$ & $r_{1/2}$ & $V_{1/2}$ & log $g_{\rm bar}$ &
  log $g_{\rm tot}$ & $\sigma_{\rm iMOND}$ & $\sigma_{\rm eMOND}$ &
  log $g_{\rm in}$ & log $g_{\rm ex}$ \\
     & ($10^5\Msun$) & (pc) & ($\kms$) & ($\acc$)  & ($\acc$) & ($\kms$) & ($\kms$) & ($\acc$) & ($\acc$) \\
  \hline
  \endfirsthead

  \hline
  \hline
   Gal. Name & $\Mstr$ & $r_{1/2}$ & $V_{1/2}$ & log $g_{\rm bar}$ &
  log $g_{\rm tot}$ & $\sigma_{\rm iMOND}$ & $\sigma_{\rm eMOND}$ &
  log $g_{\rm in}$ & log $g_{\rm ex}$ \\
  & ($10^5\Msun$) & (pc) & ($\kms$) & ($\acc$)  & ($\acc$) & ($\kms$) & ($\kms$) & ($\acc$) & ($\acc$) \\
  \hline
  \endhead
 
  \underline{{\bf MW satellites}} \\
     For  & $  245  _{-   69} ^{+   96}$ & $ 949  _{- 100} ^{+ 106}$ & $  20.2  _{-   2.8} ^{+   2.8}$ & $ -11.7  _{-   0.1} ^{+   0.1}$ & $ -10.9  _{-   0.1} ^{+   0.1}$ & $  11.8$ & $  20.6   $  & $ -13.8$ & $ -14.0 $ \\
    LeoI  & $   45  _{-   13} ^{+   19}$ & $ 334  _{-  34} ^{+  36}$ & $  15.7  _{-   2.9} ^{+   3.1}$ & $ -11.5  _{-   0.1} ^{+   0.1}$ & $ -10.6  _{-   0.2} ^{+   0.2}$ & $   7.9$ & $  20.5   $  & $ -13.7$ & $ -14.2 $ \\
     Scl  & $   39  _{-   15} ^{+   25}$ & $ 376  _{-  58} ^{+  58}$ & $  15.7  _{-   2.6} ^{+   2.8}$ & $ -11.7  _{-   0.2} ^{+   0.2}$ & $ -10.7  _{-   0.2} ^{+   0.2}$ & $   7.4$ & $   9.8   $  & $ -13.8$ & $ -13.7 $ \\
   LeoII  & $   12  _{-    3} ^{+    4.4}$ & $ 235  _{-  56} ^{+  56}$ & $  11.3  _{-   1.8} ^{+   1.8}$ & $ -11.8  _{-   0.2} ^{+   0.3}$ & $ -10.8  _{-   0.2} ^{+   0.2}$ & $   5.5$ & $  11.4   $  & $ -13.9$ & $ -14.2 $ \\
    SexI  & $    7.0  _{-    3} ^{+    4.3}$ & $ 926  _{-  56} ^{+  61}$ & $  13.6  _{-   2.7} ^{+   2.8}$ & $ -13.2  _{-   0.2} ^{+   0.2}$ & $ -11.2  _{-   0.2} ^{+   0.2}$ & $   4.8$ & $   2.7   $  & $ -14.6$ & $ -13.7 $ \\
     Car  & $    3.8  _{-    1.4} ^{+    2.3}$ & $ 334  _{-  51} ^{+  54}$ & $  11.2  _{-   2.4} ^{+   2.6}$ & $ -12.6  _{-   0.2} ^{+   0.2}$ & $ -10.9  _{-   0.2} ^{+   0.2}$ & $   4.2$ & $   3.6   $  & $ -14.3$ & $ -13.8 $ \\
     Dra  & $    5.1  _{-    1.2} ^{+    1.5}$ & $ 294  _{-  24} ^{+  25}$ & $  15.6  _{-   2.8} ^{+   2.8}$ & $ -12.4  _{-   0.1} ^{+   0.1}$ & $ -10.6  _{-   0.2} ^{+   0.2}$ & $   4.5$ & $   3.8   $  & $ -14.2$ & $ -13.7 $ \\
     Umi  & $    5.3  _{-    2.0} ^{+    3.3}$ & $ 584  _{-  62} ^{+  63}$ & $  16.3  _{-   2.8} ^{+   2.9}$ & $ -13.0  _{-   0.2} ^{+   0.2}$ & $ -10.8  _{-   0.2} ^{+   0.1}$ & $   4.5$ & $   2.8   $  & $ -14.5$ & $ -13.7 $ \\
 CanVenI  & $    3.7  _{-    0.8} ^{+    0.9}$ & $ 751  _{-  47} ^{+  49}$ & $  13.1  _{-   1.7} ^{+   1.7}$ & $ -13.3  _{-   0.1} ^{+   0.1}$ & $ -11.1  _{-   0.1} ^{+   0.1}$ & $   4.1$ & $   3.4   $  & $ -14.7$ & $ -14.1 $ \\
   CraII  & $    2.6  _{-    0.3} ^{+    0.4}$ & $1332  _{- 100} ^{+ 109}$ & $   4.8  _{-   0.8} ^{+   0.8}$ & $ -14.0  _{-   0.1} ^{+   0.1}$ & $ -12.3  _{-   0.1} ^{+   0.1}$ & $   3.8$ & $   1.6   $  & $ -15.0$ & $ -13.9 $ \\
     Her  & $  0.60  _{-  0.18 } ^{+  0.23}$ & $ 443  _{-  71} ^{+  99}$ & $   6.3  _{-   1.7} ^{+   1.8}$ & $ -13.7  _{-   0.2} ^{+   0.2}$ & $ -11.5  _{-   0.3} ^{+   0.2}$ & $   2.3$ & $   1.1   $  & $ -14.9$ & $ -13.9 $ \\
    BooI  & $  0.46  _{-  0.09 } ^{+  0.11}$ & $ 325  _{-  28} ^{+  29}$ & $   4.0  _{-   1.5} ^{+   1.1}$ & $ -13.5  _{-   0.1} ^{+   0.1}$ & $ -11.8  _{-   0.4} ^{+   0.2}$ & $   2.2$ & $   0.8   $  & $ -14.8$ & $ -13.6 $ \\
   LeoIV  & $  0.29  _{-  0.09 } ^{+  0.14}$ & $ 275  _{-  49} ^{+  50}$ & $   5.7  _{-   2.7} ^{+   3.1}$ & $ -13.6  _{-   0.2} ^{+   0.2}$ & $ -11.4  _{-   0.6} ^{+   0.4}$ & $   2.0$ & $   1.1   $  & $ -14.9$ & $ -14.0 $ \\
    UMaI  & $  0.22  _{-  0.06 } ^{+  0.08}$ & $ 423  _{-  65} ^{+  68}$ & $  13.1  _{-   2.3} ^{+   2.4}$ & $ -14.1  _{-   0.2} ^{+   0.2}$ & $ -10.9  _{-   0.2} ^{+   0.2}$ & $   1.8$ & $   0.6   $  & $ -15.1$ & $ -13.8 $ \\
    LeoV  & $  0.17  _{-  0.06 } ^{+  0.08}$ & $ 179  _{-  42} ^{+  43}$ & $   4.7  _{-   2.8} ^{+   2.6}$ & $ -13.4  _{-   0.2} ^{+   0.3}$ & $ -11.4  _{-   0.8} ^{+   0.4}$ & $   1.7$ & $   1.1   $  & $ -14.8$ & $ -14.1 $ \\
   PisII  & $  0.14  _{-  0.07 } ^{+  0.13}$ & $  77  _{-  17} ^{+  22}$ & $   9.4  _{-   5.1} ^{+   4.4}$ & $ -12.8  _{-   0.2} ^{+   0.2}$ & $ -10.4  _{-   0.7} ^{+   0.4}$ & $   1.6$ & $   1.5   $  & $ -14.5$ & $ -14.1 $ \\
CanVeniI  & $  0.13  _{-  0.05 } ^{+  0.08}$ & $  99  _{-  18} ^{+  19}$ & $   7.8  _{-   1.9} ^{+   2.1}$ & $ -13.0  _{-   0.3} ^{+   0.3}$ & $ -10.7  _{-   0.3} ^{+   0.2}$ & $   1.6$ & $   1.2   $  & $ -14.6$ & $ -14.0 $ \\
   HydII  & $  0.13  _{-  0.03 } ^{+  0.05}$ & $  89  _{-  13} ^{+  16}$ & $   7.1  _{-   1.2} ^{+   5.5}$ & $ -13.0  _{-   0.2} ^{+   0.2}$ & $ -10.7  _{-   0.2} ^{+   0.5}$ & $   1.6$ & $   1.1   $  & $ -14.6$ & $ -13.9 $ \\
   UMaII  & $  0.065  _{-  0.03 } ^{+  0.05}$ & $ 196  _{-  27} ^{+  31}$ & $  11.4  _{-   2.5} ^{+   3.0}$ & $ -13.9  _{-   0.2} ^{+   0.2}$ & $ -10.7  _{-   0.2} ^{+   0.2}$ & $   1.3$ & $   0.3   $  & $ -15.0$ & $ -13.4 $ \\
  ComBer  & $  0.060  _{-  0.02 } ^{+  0.04}$ & $ 101  _{-  13} ^{+  14}$ & $   7.9  _{-   1.6} ^{+   1.7}$ & $ -13.4  _{-   0.2} ^{+   0.2}$ & $ -10.7  _{-   0.2} ^{+   0.2}$ & $   1.3$ & $   0.4   $  & $ -14.8$ & $ -13.5 $ \\
    Tuc2  & $  0.045  _{-  0.01 } ^{+  0.012}$ & $ 222  _{-  33} ^{+  41}$ & $  14.4  _{-   6.6} ^{+   5.3}$ & $ -14.2  _{-   0.1} ^{+   0.1}$ & $ -10.5  _{-   0.5} ^{+   0.3}$ & $   1.2$ & $   0.3   $  & $ -15.2$ & $ -13.5 $ \\
    Hor1  & $  0.031  _{-  0.006 } ^{+  0.008}$ & $  41  _{-   6} ^{+   7}$ & $   8.0  _{-   3.9} ^{+   2.2}$ & $ -12.9  _{-   0.1} ^{+   0.1}$ & $ -10.3  _{-   0.6} ^{+   0.2}$ & $   1.1$ & $   0.7   $  & $ -14.5$ & $ -13.7 $ \\
    Gru1  & $  0.031  _{-  0.009 } ^{+  0.013}$ & $  84  _{-  21} ^{+  38}$ & $   6.5  _{-   3.3} ^{+   3.0}$ & $ -13.5  _{-   0.3} ^{+   0.3}$ & $ -10.8  _{-   0.6} ^{+   0.4}$ & $   1.1$ & $   0.6   $  & $ -14.8$ & $ -13.9 $ \\
   DraII  & $  0.020  _{-  0.011 } ^{+  0.025}$ & $  25  _{-   8} ^{+  10}$ & $   5.3  _{-   3.0} ^{+   3.5}$ & $ -12.7  _{-   0.4} ^{+   0.4}$ & $ -10.4  _{-   0.7} ^{+   0.5}$ & $   1.0$ & $   0.4   $  & $ -14.4$ & $ -13.2 $ \\
   BooII  & $  0.016  _{-  0.009 } ^{+  0.021}$ & $  68  _{-  22} ^{+  23}$ & $  19.3  _{-  10.8} ^{+  12.7}$ & $ -13.6  _{-   0.5} ^{+   0.5}$ & $  -9.7  _{-   0.7} ^{+   0.5}$ & $   1.0$ & $   0.3   $  & $ -14.9$ & $ -13.4 $ \\
    Ret2  & $  0.016  _{-  0.003 } ^{+  0.004}$ & $  43  _{-   4} ^{+   5}$ & $   5.3  _{-   2.4} ^{+   1.4}$ & $ -13.2  _{-   0.1} ^{+   0.1}$ & $ -10.7  _{-   0.5} ^{+   0.2}$ & $   1.0$ & $   0.3   $  & $ -14.7$ & $ -13.3 $ \\
   Will1  & $  0.016  _{-  0.008 } ^{+  0.018}$ & $  33  _{-   8} ^{+   9}$ & $   7.1  _{-   3.3} ^{+   2.7}$ & $ -13.0  _{-   0.3} ^{+   0.3}$ & $ -10.3  _{-   0.6} ^{+   0.3}$ & $   1.0$ & $   0.4   $  & $ -14.6$ & $ -13.4 $ \\
   SegII  & $  0.014  _{-  0.004 } ^{+  0.005}$ & $  46  _{-   3} ^{+   3}$ & $   3.8  _{-   0.6} ^{+   0.7}$ & $ -13.3  _{-   0.1} ^{+   0.1}$ & $ -11.0  _{-   0.2} ^{+   0.1}$ & $   0.9$ & $   0.3   $  & $ -14.8$ & $ -13.4 $ \\
   TriII  & $  0.0071 _{-  0.0027 } ^{+  0.0045}$ & $  46  _{-  11} ^{+  12}$ & $   6.6  _{-   1.1} ^{+   5.1}$ & $ -13.6  _{-   0.3} ^{+   0.3}$ & $ -10.5  _{-   0.2} ^{+   0.5}$ & $   0.8$ & $   0.2   $  & $ -14.9$ & $ -13.4 $ \\
    SegI  & $  0.0054  _{-  0.0029 } ^{+  0.0063}$ & $  40  _{-   7} ^{+  11}$ & $   6.6  _{-   1.5} ^{+   1.7}$ & $ -13.6  _{-   0.4} ^{+   0.4}$ & $ -10.5  _{-   0.2} ^{+   0.2}$ & $   0.7$ & $   0.2   $  & $ -14.9$ & $ -13.2 $ \\
 & \\
\underline{{\bf M31 satellites}} \\
    N205  & $ 4650  _{-  650} ^{+  760}$ & $ 786  _{-  40} ^{+  41}$ & $  59.9  _{-  10.5} ^{+  11.0}$ & $ -10.3  _{-   0.1} ^{+   0.1}$ & $  -9.8  _{-   0.2} ^{+   0.1}$ & $  24.6$ & $  50.1   $  & $ -13.1$ & $ -13.4$ \\ 
     M32  & $ 4760  _{- 1010} ^{+ 1260}$ & $ 146  _{-  20} ^{+  22}$ & $  86.7  _{-   9.1} ^{+   8.3}$ & $  -8.8  _{-   0.1} ^{+   0.1}$ & $  -8.8  _{-   0.1} ^{+   0.1}$ & $  24.8$ & $  86.5   $  & $ -12.4$ & $ -13.1$ \\ 
    N185  & $  680  _{-   100} ^{+  118}$ & $ 358  _{-  14} ^{+  16}$ & $  41.5  _{-   4.6} ^{+   4.3}$ & $ -10.4  _{-   0.1} ^{+   0.1}$ & $  -9.8  _{-   0.1} ^{+   0.1}$ & $  15.2$ & $  59.9   $  & $ -13.2$ & $ -14.0$ \\ 
    N147  & $  990  _{-  150} ^{+  164}$ & $ 532  _{-  21} ^{+  22}$ & $  27.5  _{-   3.3} ^{+   3.3}$ & $ -10.6  _{-   0.1} ^{+   0.1}$ & $ -10.3  _{-   0.1} ^{+   0.1}$ & $  16.7$ & $  51.8   $  & $ -13.3$ & $ -13.9$ \\ 
    AVII  & $  150  _{-   39} ^{+   52}$ & $1034  _{-  54} ^{+  58}$ & $  22.5  _{-   2.8} ^{+   2.8}$ & $ -12.0  _{-   0.1} ^{+   0.1}$ & $ -10.8  _{-   0.1} ^{+   0.1}$ & $  10.4$ & $  17.8   $  & $ -14.0$ & $ -14.1$ \\ 
     AII  & $   40.  _{-    8} ^{+    9}$ & $1340  _{-  44} ^{+  45}$ & $  13.4  _{-   2.3} ^{+   2.4}$ & $ -12.8  _{-   0.1} ^{+   0.1}$ & $ -11.4  _{-   0.2} ^{+   0.1}$ & $   7.5$ & $   7.4   $  & $ -14.4$ & $ -14.0$ \\ 
      AI  & $   44.  _{-    9} ^{+   10}$ & $1125  _{-  46} ^{+  47}$ & $  17.5  _{-   3.6} ^{+   3.9}$ & $ -12.6  _{-   0.1} ^{+   0.1}$ & $ -11.1  _{-   0.2} ^{+   0.2}$ & $   7.7$ & $   4.8   $  & $ -14.3$ & $ -13.5$ \\ 
     AVI  & $   50.  _{-    10} ^{+   12}$ & $ 698  _{-  64} ^{+  64}$ & $  21.3  _{-   3.0} ^{+   3.3}$ & $ -12.1  _{-   0.1} ^{+   0.1}$ & $ -10.7  _{-   0.1} ^{+   0.1}$ & $   7.9$ & $  13.9   $  & $ -14.1$ & $ -14.2$ \\ 
  AXXIII  & $   11.  _{-    2} ^{+    3}$ & $1608  _{- 145} ^{+ 155}$ & $  12.1  _{-   2.0} ^{+   2.2}$ & $ -13.5  _{-   0.1} ^{+   0.1}$ & $ -11.5  _{-   0.2} ^{+   0.1}$ & $   5.4$ & $   2.9   $  & $ -14.8$ & $ -13.9$ \\ 
    AIII  & $   10.  _{-    2} ^{+    3}$ & $ 581  _{-  61} ^{+  62}$ & $  15.9  _{-   2.8} ^{+   3.0}$ & $ -12.7  _{-   0.1} ^{+   0.1}$ & $ -10.8  _{-   0.2} ^{+   0.2}$ & $   5.3$ & $   3.7   $  & $ -14.3$ & $ -13.6$ \\ 
    LGS3  & $    9.6  _{-    1} ^{+    1.6}$ & $ 625  _{-  61} ^{+  64}$ & $  13.9  _{-   5.3} ^{+   8.9}$ & $ -12.8  _{-   0.1} ^{+   0.1}$ & $ -11.0  _{-   0.4} ^{+   0.4}$ & $   5.3$ & $   6.5   $  & $ -14.4$ & $ -14.2$ \\ 
    AXXI  & $    5.5  _{-    1.4} ^{+    1.9}$ & $1324  _{- 142} ^{+ 247}$ & $   7.8  _{-   2.0} ^{+   2.2}$ & $ -13.7  _{-   0.2} ^{+   0.2}$ & $ -11.8  _{-   0.3} ^{+   0.2}$ & $   4.6$ & $   2.4   $  & $ -14.8$ & $ -13.9$ \\ 
    AXXV  & $    6.3  _{-    1.6} ^{+    1.8}$ & $ 864  _{-  88} ^{+ 125}$ & $   5.1  _{-   1.9} ^{+   2.1}$ & $ -13.2  _{-   0.2} ^{+   0.1}$ & $ -12.0  _{-   0.4} ^{+   0.3}$ & $   4.7$ & $   2.6   $  & $ -14.6$ & $ -13.7$ \\ 
      AV  & $    5.1  _{-    1.0} ^{+    1.3}$ & $ 484  _{-  38} ^{+  58}$ & $  18.0  _{-   2.6} ^{+   2.7}$ & $ -12.8  _{-   0.1} ^{+   0.1}$ & $ -10.7  _{-   0.1} ^{+   0.1}$ & $   4.5$ & $   3.4   $  & $ -14.4$ & $ -13.8$ \\ 
     AXV  & $    2.1  _{-    0.6} ^{+    0.8}$ & $ 320  _{-  32} ^{+  43}$ & $   6.8  _{-   2.4} ^{+   2.6}$ & $ -12.9  _{-   0.1} ^{+   0.1}$ & $ -11.3  _{-   0.4} ^{+   0.3}$ & $   3.6$ & $   3.4   $  & $ -14.4$ & $ -14.0$ \\ 
    AXIX  & $   12.  _{-    4} ^{+    5}$ & $4374  _{- 842} ^{+1069}$ & $   8.0  _{-   2.5} ^{+   3.0}$ & $ -14.4  _{-   0.2} ^{+   0.2}$ & $ -12.3  _{-   0.3} ^{+   0.3}$ & $   5.6$ & $   1.8   $  & $ -15.2$ & $ -13.8$ \\ 
    AXIV  & $    3.3  _{-    1.2} ^{+    1.5}$ & $ 431  _{-  91} ^{+  77}$ & $   9.0  _{-   1.9} ^{+   2.0}$ & $ -12.9  _{-   0.2} ^{+   0.2}$ & $ -11.2  _{-   0.2} ^{+   0.2}$ & $   4.0$ & $   3.6   $  & $ -14.4$ & $ -14.0$ \\ 
   AXXIX  & $    2.7  _{-    0.9} ^{+    1.5}$ & $ 481  _{-  73} ^{+  78}$ & $   9.7  _{-   2.2} ^{+   2.3}$ & $ -13.1  _{-   0.2} ^{+   0.2}$ & $ -11.2  _{-   0.2} ^{+   0.2}$ & $   3.8$ & $   3.3   $  & $ -14.5$ & $ -14.0$ \\ 
     AIX  & $    4.3  _{-    1.1} ^{+    1.5}$ & $ 596  _{-  64} ^{+  90}$ & $  18.8  _{-   4.0} ^{+   4.1}$ & $ -13.1  _{-   0.2} ^{+   0.2}$ & $ -10.7  _{-   0.2} ^{+   0.2}$ & $   4.3$ & $   1.7   $  & $ -14.5$ & $ -13.4$ \\ 
    AXXX  & $    2.2  _{-    0.6} ^{+    0.7}$ & $ 385  _{-  59} ^{+  62}$ & $  20.3  _{-   8.2} ^{+  13.2}$ & $ -13.0  _{-   0.2} ^{+   0.2}$ & $ -10.5  _{-   0.5} ^{+   0.4}$ & $   3.6$ & $   2.9   $  & $ -14.5$ & $ -13.9$ \\ 
  AXXVII  & $  2.0  _{-  0.8 } ^{+  1.8}$ & $ 580  _{- 101} ^{+ 102}$ & $  24.7  _{-   7.4} ^{+   6.9}$ & $ -13.4  _{-   0.3} ^{+   0.3}$ & $ -10.5  _{-   0.3} ^{+   0.2}$ & $   3.1$ & $   1.3   $  & $ -14.8$ & $ -13.6 $ \\
   AXVII  & $  1.7  _{-  0.4 } ^{+  0.6}$ & q$ 398  _{-  85} ^{+  87}$ & $   5.3  _{-   3.1} ^{+   3.2}$ & $ -13.1  _{-   0.2} ^{+   0.2}$ & $ -11.6  _{-   0.8} ^{+   0.4}$ & $   3.0$ & $   1.4   $  & $ -14.7$ & $ -13.6 $ \\
      AX  & $  1.3  _{-  0.3 } ^{+  0.5}$ & $ 287  _{-  56} ^{+ 102}$ & $  10.9  _{-   2.6} ^{+   2.9}$ & $ -13.0  _{-   0.3} ^{+   0.2}$ & $ -10.9  _{-   0.3} ^{+   0.2}$ & $   2.8$ & $   2.0   $  & $ -14.6$ & $ -13.9 $ \\
    AXVI  & $  1.1  _{-  0.3 } ^{+  0.5}$ & $ 185  _{-  22} ^{+  25}$ & $   7.0  _{-   4.1} ^{+   4.8}$ & $ -12.6  _{-   0.2} ^{+   0.2}$ & $ -11.1  _{-   0.8} ^{+   0.5}$ & $   2.7$ & $   3.5   $  & $ -14.4$ & $ -14.3 $ \\
    AXII  & $  0.85  _{-  0.35 } ^{+  0.62}$ & $ 369  _{-  78} ^{+  92}$ & $   6.1  _{-   3.6} ^{+   3.9}$ & $ -13.4  _{-   0.3} ^{+   0.3}$ & $ -11.5  _{-   0.8} ^{+   0.4}$ & $   2.6$ & $   1.6   $  & $ -14.8$ & $ -14.0 $ \\
   AXIII  & $  0.73  _{-  0.23 } ^{+  0.33}$ & $ 263  _{-  97} ^{+ 128}$ & $   9.9  _{-   3.5} ^{+   3.8}$ & $ -13.1  _{-   0.4} ^{+   0.4}$ & $ -10.9  _{-   0.4} ^{+   0.3}$ & $   2.5$ & $   1.5   $  & $ -14.6$ & $ -13.9 $ \\
   AXXII  & $  0.73  _{-  0.23 } ^{+  0.33}$ & $ 323  _{-  74} ^{+ 105}$ & $   4.9  _{-   2.7} ^{+   2.5}$ & $ -13.3  _{-   0.3} ^{+   0.3}$ & $ -11.6  _{-   0.7} ^{+   0.4}$ & $   2.5$ & $   2.0   $  & $ -14.7$ & $ -14.2 $ \\
     AXX  & $  0.48  _{-  0.16 } ^{+  0.22}$ & $ 116  _{-  31} ^{+  57}$ & $  12.1  _{-   5.8} ^{+   4.7}$ & $ -12.6  _{-   0.4} ^{+   0.3}$ & $ -10.4  _{-   0.6} ^{+   0.3}$ & $   2.2$ & $   1.8   $  & $ -14.4$ & $ -13.9 $ \\
     AXI  & $  0.46  _{-  0.15 } ^{+  0.22}$ & $ 171  _{-  58} ^{+  57}$ & $  13.1  _{-   6.3} ^{+   5.1}$ & $ -12.9  _{-   0.3} ^{+   0.4}$ & $ -10.5  _{-   0.6} ^{+   0.3}$ & $   2.2$ & $   1.4   $  & $ -14.6$ & $ -13.8 $ \\
   AXXVI  & $  0.31  _{-  0.15 } ^{+  0.20}$ & $ 303  _{- 141} ^{+ 182}$ & $  14.4  _{-   4.7} ^{+   4.5}$ & $ -13.6  _{-   0.5} ^{+   0.6}$ & $ -10.7  _{-   0.4} ^{+   0.4}$ & $   2.0$ & $   0.8   $  & $ -14.9$ & $ -13.8 $ \\
 & \\
\underline{{\bf LG field dwarfs}} \\
   N6822  & $  830  _{-  170} ^{+  200}$ & $ 638_{  48} ^{  50}$ & $  40.0  _{-   4.6} ^{+   4.5}$ & $ -10.9  _{-   0.1} ^{+   0.1}$ & $ -10.1  _{-   0.1} ^{+   0.1}$ & $  16.0$ & $  80.4   $  & $ -13.4$ & $ -14.5 $ \\
  IC1613  & $ 1020  _{-  170} ^{+  190}$ & $1383_{ 135} ^{ 144}$ & $  18.7  _{-   2.5} ^{+   2.5}$ & $ -11.4  _{-   0.1} ^{+   0.1}$ & $ -11.1  _{-   0.1} ^{+   0.1}$ & $  16.9$ & $  62.3   $  & $ -13.7$ & $ -14.5 $ \\
     WLM  & $  380  _{-   55} ^{+   65}$ & $2090_{ 139} ^{ 149}$ & $  29.4  _{-   3.5} ^{+   3.3}$ & $ -12.2  _{-   0.1} ^{+   0.1}$ & $ -10.9  _{-   0.1} ^{+   0.1}$ & $  13.2$ & $  39.4   $  & $ -14.1$ & $ -14.7 $ \\
 UGC4879  & $   58  _{-   11} ^{+   13}$ & $ 217_{  22} ^{  21}$ & $  16.6  _{-   2.6} ^{+   2.9}$ & $ -11.1  _{-   0.1} ^{+   0.1}$ & $ -10.4  _{-   0.2} ^{+   0.1}$ & $   8.2$ & $  63.7   $  & $ -13.5$ & $ -14.9 $ \\
     Peg  & $   66  _{-   13} ^{+   16}$ & $ 928_{  64} ^{  66}$ & $  21.2  _{-   2.8} ^{+   3.0}$ & $ -12.3  _{-   0.1} ^{+   0.1}$ & $ -10.8  _{-   0.1} ^{+   0.1}$ & $   8.5$ & $  18.4   $  & $ -14.1$ & $ -14.4 $ \\
    LeoA  & $   30  _{-    6} ^{+    8}$ & $ 471_{  42} ^{  45}$ & $  11.5  _{-   2.4} ^{+   2.7}$ & $ -12.0  _{-   0.1} ^{+   0.1}$ & $ -11.0  _{-   0.2} ^{+   0.2}$ & $   7.0$ & $  23.6   $  & $ -14.0$ & $ -14.7 $ \\
     Cet  & $   45  _{-    9} ^{+   11}$ & $ 936_{  41} ^{  43}$ & $  14.3  _{-   2.2} ^{+   2.3}$ & $ -12.4  _{-   0.1} ^{+   0.1}$ & $ -11.2  _{-   0.1} ^{+   0.1}$ & $   7.7$ & $  18.1   $  & $ -14.2$ & $ -14.6 $ \\
     Aqu  & $   16  _{-    2} ^{+    3}$ & $ 611_{  27} ^{  28}$ & $  13.5  _{-   3.0} ^{+   3.6}$ & $ -12.5  _{-   0.1} ^{+   0.1}$ & $ -11.0  _{-   0.2} ^{+   0.2}$ & $   5.9$ & $  17.3   $  & $ -14.3$ & $ -14.8 $ \\
     Tuc  & $    8.9  _{-    1.9} ^{+    2.3}$ & $ 378_{  21} ^{  23}$ & $  27.4  _{-   6.2} ^{+   7.3}$ & $ -12.4  _{-   0.1} ^{+   0.1}$ & $ -10.2  _{-   0.2} ^{+   0.2}$ & $   5.2$ & $  15.2   $  & $ -14.2$ & $ -14.8 $ \\
  AXVIII  & $    8.0  _{-    1.6} ^{+    1.9}$ & $ 433_{  31} ^{  32}$ & $  16.6  _{-   4.1} ^{+   4.5}$ & $ -12.5  _{-   0.1} ^{+   0.1}$ & $ -10.7  _{-   0.2} ^{+   0.2}$ & $   5.0$ & $   9.2   $  & $ -14.2$ & $ -14.4 $ \\
 AXXVIII  & $    4.1  _{-    1.7} ^{+    5.2}$ & $ 294_{  65} ^{  82}$ & $  11.5  _{-   3.9} ^{+   5.2}$ & $ -12.5  _{-   0.3} ^{+   0.4}$ & $ -10.8  _{-   0.4} ^{+   0.3}$ & $   4.2$ & $   7.2   $  & $ -14.2$ & $ -14.3 $ \\
    LeoT  & $    1.1  _{-    0.4} ^{+    0.6}$ & $ 160_{  12} ^{  12}$ & $  12.8  _{-   3.0} ^{+   3.2}$ & $ -12.5  _{-   0.2} ^{+   0.2}$ & $ -10.5  _{-   0.2} ^{+   0.2}$ & $   3.0$ & $   5.6   $  & $ -14.2$ & $ -14.4 $ \\
   EriII  & $    0.9  _{-    0.2} ^{+    0.2}$ & $ 325_{  23} ^{  23}$ & $  11.9  _{-   2.0} ^{+   2.3}$ & $ -13.2  _{-   0.1} ^{+   0.1}$ & $ -10.8  _{-   0.2} ^{+   0.2}$ & $   2.9$ & $   3.5   $  & $ -14.6$ & $ -14.4 $ \\

  \hline

  \label{TabData2}
\end{longtable}

\onecolumn
\def\arraystretch{1.3}
\begin{longtable}{{l} *{4}{c} }
  \caption{Derived parameters for progenitors of MW and M31 satellites}\\
  
  \hline
  \hline
  Progenitor of  &  $\Mstr$ & $3D r_{1/2}$ &  $V_{1/2}$ & $\mu_{\rm L}$  \\
  & ($10^5\Msun$) & (pc) & ($\kms$) & \\ 
  \hline
  \endfirsthead

  {{\bfseries \tablename\ \thetable{} -- (continued)}} \\
  \hline
  \hline
  Progenitor  &  $\Mstr$ & $3D r_{1/2}$ &  $V_{1/2}$ & $\mu_{\rm L}$  \\
  & ($10^5\Msun$) & (pc) & ($\kms$) & \\
  \hline
  \endhead

  \underline{{\bf MW satellites}} \\
     For &     300. &   864 &  25.8 &  0.81 \\
    LeoI &     49.4 &   334 &  15.7 &  1.00 \\
     Scl &     39.1 &   371 &  16.2 &  0.99 \\
   LeoII &     12.4 &   225 &  12.4 &  0.95 \\
    SexI &     11.5 &   813 &  20.8 &  0.60 \\
     Car &      4.4 &   307 &  13.8 &  0.85 \\
     Dra &      5.1 &   294 &  15.6 &  1.00 \\
     Umi &      5.6 &   563 &  17.7 &  0.95 \\
 CanVenI &      5.4 &   665 &  18.9 &  0.68 \\
   CraII &    264 &  1819 &  18.1 &  0.01 \\
     Her &      4.9 &   426 &  15.9 &  0.12 \\
    BooI &     46.3 &   443 &  15.7 &  0.01 \\
   LeoIV &      1.28 &   250 &  12.2 &  0.23 \\
    UMaI &      0.23 &   411 &  14.1 &  0.96 \\
    LeoV &      2.25 &   183 &  10.8 &  0.08 \\
   PisII &      0.137 &    77 &   9.4 &  1.00 \\
CanVeniI &      0.127 &    99 &   8.0 &  1.00 \\
   HydII &      0.127 &    84 &   7.0 &  0.92 \\
   UMaII &      0.065 &   196 &  11.6 &  1.00 \\
  ComBer &      0.060 &   101 &   8.0 &  1.00 \\
    Tuc2 &      0.045 &   219 &  14.9 &  1.00 \\
    Hor1 &      0.031 &    40 &   8.5 &  0.99 \\
    Gru1 &      0.039 &    74 &   6.5 &  0.79 \\
   DraII &      0.020 &    25 &   5.0 &  1.00 \\
   BooII &      0.016 &    67 &  18.2 &  1.00 \\
    Ret2 &      0.016 &    42 &   5.6 &  0.99 \\
   Will1 &      0.016 &    33 &   7.4 &  0.98 \\
   SegII &      0.017 &    41 &   4.9 &  0.81 \\
   TriII &      0.007 &    45 &   5.9 &  0.99 \\
    SegI &      0.005 &    39 &   6.8 &  0.99 \\
& \\ 
\underline{{\bf M31 satellites}} \\
    N205 &   4650 &   786 &  59.9 &  1.00 \\
     M32 &   5070 &   146 &  86.7 &  1.00 \\
    N185 &    676 &   358 &  41.5 &  1.00 \\
    N147 &    988 &   532 &  27.5 &  1.00 \\
    AVII &    161 &   977 &  25.7 &  0.92 \\
     AII &    145 &  1203 &  27.3 &  0.27 \\
      AI &     60.0 &  1004 &  24.3 &  0.72 \\
     AVI &     52.8 &   698 &  21.3 &  1.00 \\
  AXXIII &     64.2 &  1495 &  27.5 &  0.18 \\
    AIII &     11.1 &   551 &  18.0 &  0.93 \\
    LGS3 &     12.2 &   565 &  18.2 &  0.79 \\
    AXXI &    593 &  1808 &  29.3 &  0.01 \\
    AXXV &    681 &  1180 &  19.4 &  0.01 \\
      AV &      5.12 &   484 &  18.0 &  1.00 \\
     AXV &      7.42 &   285 &  13.6 &  0.30 \\
    AXIX &   1336 &  5975 &  30.4 &  0.01 \\
    AXIV &      7.28 &   377 &  15.3 &  0.49 \\
   AXXIX &      5.35 &   421 &  15.7 &  0.54 \\
     AIX &      4.59 &   596 &  18.8 &  1.00 \\
    AXXX &      2.34 &   385 &  20.3 &  1.00 \\
  AXXVII &      1.96 &   578 &  25.6 &  1.03 \\
   AXVII &    168 &   539 &  19.0 &  0.01 \\
      AX &      1.40 &   270 &  12.7 &  0.90 \\
    AXVI &      1.91 &   162 &  10.2 &  0.60 \\
    AXII &     87.4 &   507 &  17.0 &  0.01 \\
   AXIII &      0.83 &   242 &  11.9 &  0.88 \\
   AXXII &     47.5 &   411 &  17.2 &  0.02 \\
     AXX &      0.474 &   115 &  12.3 &  1.01 \\
     AXI &      0.466 &   170 &  13.2 &  0.99 \\
   AXXVI &      0.316 &   295 &  14.9 &  0.99 \\

  \label{TabProg}
\end{longtable}

\label{lastpage}
\end{document}